\documentclass[a4paper,11pt]{article}
\usepackage[T1]{fontenc}
\usepackage[scr=boondox]{mathalpha}
\usepackage{amsmath,calligra}
\usepackage{mathrsfs}
\usepackage{comment}

\usepackage{jheppub}
\usepackage{booktabs,threeparttable,makecell,array}
\usepackage[table]{xcolor}
\usepackage{pifont}
\usepackage{arydshln} 
\usepackage[normalem]{ulem}
\usepackage{tabularx}

\newcommand{\cmark}{\ding{51}}
\newcommand{\xmark}{\ding{55}}

\definecolor{Hbar}{rgb}{0.95,0.98,0.94}

\newcolumntype{L}{>{\raggedright\arraybackslash}p{3.8cm}}
\newcolumntype{C}{>{\centering\arraybackslash}c}
\newcolumntype{F}{>{\centering\arraybackslash}p{1.2cm}}
\newcolumntype{G}{>{\columncolor{Hbar}\centering\arraybackslash}c}
\newcolumntype{R}{>{\centering\arraybackslash}c}

\renewcommand{\arraystretch}{1.18}
\definecolor{MixRow}{rgb}{0.96,0.98,0.95}

\usepackage{hyperref}
\usepackage{cleveref}
\usepackage{physics}
\usepackage{tikz}
\usepackage{multirow}
\usepackage{rotating}
\usepackage{upgreek}
\usepackage{graphicx}
\usepackage{etoolbox}
\usepackage{siunitx}
\usepackage{xcolor}
\usepackage{subcaption}
\usepackage{enumitem}
  \newcommand{\llangle}{\langle\!\!\langle}
\newcommand{\rrangle}{\rangle\!\!\rangle}
\usepackage{extarrows} 
\usepackage{amsmath}
\usepackage{dcolumn}
\usepackage{simpler-wick}
\usepackage{kantlipsum}
\usepackage{float}
\usepackage{rotating}

\usepackage{etoolbox}
\usepackage{extarrows} 
\usepackage{bm}
\usepackage{wasysym}
\usepackage{verbatim}

\usepackage{verbatim}
\usepackage{color}
\usepackage{mathtools,slashed}

\setcitestyle{square}
\usepackage[scr=boondox]{mathalpha}
\usepackage{xcolor}
\usepackage{braket}
\usepackage{upgreek}
\usepackage{cancel}
\usepackage{amsmath}
\usepackage{amssymb}
\usepackage{mathtools}
\usepackage{slashed}
\newcommand{\p}{\partial}

\newcommand{\bx}{\boldsymbol{x}}
\newcommand{\be}{\boldsymbol{e}}

\newcommand{\bq}{\boldsymbol{q}}
\newcommand{\bp}{\boldsymbol{p}}

\crefname{equation}{}{}

\title{The Stochastic Schwinger Effect}%

\abstract{We formulate a stochastic generalisation of the Schwinger effect, extending pair production to statistically fluctuating gauge-field backgrounds. Our approach captures realistic field configurations that are transient, inhomogeneous, and stochastic, as commonly encountered in cosmological and high-energy astrophysical settings. Using the effective action formalism, we compute the vacuum decay rate and number density of charged particles, obtaining closed-form analytical expressions for both scalar and fermionic cases. To isolate the essential physics, the analysis is performed in flat spacetime and at zero temperature, providing a controlled setting in which curvature and thermal effects can be neglected. As a proof of concept, we present representative phenomenological examples relevant to astrophysical plasmas and early-Universe–motivated scenarios.}

\preprint{KCL-PH-TH/2025-41}
\author[a]{Lucas Vicente García-Consuegra}
\emailAdd{lucas.vicente\_garcia-consuegra@kcl.ac.uk}
\affiliation[a]{Department of Physics, King's College London, Strand, London, WC2R 2LS, U.K.}
\author[b]{and, Azadeh Maleknejad}
\emailAdd{azadeh.maleknejad@swansea.ac.uk}
\affiliation[b]{Centre for Quantum Fields and Gravity, Department of Physics, Swansea University,\\
Singleton Park, Swansea, SA2 8PP, U.K.}
\date{\today}%

\begin{document}
\maketitle

\section{Introduction\label{Sec:Introduction}}

One of the most profound predictions of quantum electrodynamics (QED) is the spontaneous creation of matter from the vacuum. 
Building on the pioneering analyses of Euler-Heisenberg~\cite{heisenberg1929folgerungen} and Sauter~\cite{RefWorks:sauter1931ber}, 
Schwinger~\cite{Schwinger1951} provided the first fully non-perturbative formulation of this phenomenon, 
establishing a cornerstone of strong-field QED. 
Direct evidence of static Schwinger pair production remains limited to analogue systems~\cite{Berdyugin2022,RefWorks:piñeiro2019sauterschwinger,RefWorks:martinez2016realtime}, 
as the required (classical background) field strengths lie far beyond current experimental capabilities. 
Despite this, the Schwinger mechanism, continues to attract extensive theoretical attention, 
with numerous variants examined in the literature. 
These include deterministic (classical-field) realisations such as the 
inhomogeneous~\cite{Hebenstreit:2011pm}, 
dynamically assisted~\cite{Brezin:1970xf, Kim:2007pm}, 
and thermally assisted~\cite{Hallin:1994ad,Gould:2017fve}, 
as well as the genuinely quantum Breit--Wheeler regime~\cite{Breit:1934zz}, which was recently observed by the STAR Collaboration~\cite{STAR:2021twy}. 
A concise overview of these different frameworks is provided in Table \ref{tab:processes}.

\begin{table}[H]
  \centering
  \small
  \begin{threeparttable}
    \setlength{\tabcolsep}{4pt}
    \renewcommand{\arraystretch}{1.3}
    \definecolor{detcolor}{gray}{0.97}
    \definecolor{stochcolor}{RGB}{110,204,239}
    \colorlet{stochcolorAlpha}{stochcolor!50}
    \begin{tabular}{L >{\columncolor{detcolor}}C >{\columncolor{detcolor}}C >{\columncolor{detcolor}}C >{\columncolor{stochcolorAlpha}}C C C R}
      \toprule
      & \multicolumn{4}{c}{\textbf{Classical}}
      & \textbf{Quantum} & \textbf{Thermal}
       \\
      \cmidrule(lr){2-5}\cmidrule(lr){6-6}\cmidrule(lr){7-7}
      \makecell[l]{\textbf{Production}\\[-2pt]\textbf{mechanism}}
      & \(E_{0}\)
      & \(E_{1}\cos(\omega t)\)
      & \(E(t,\bx)\)
      & \(\langle E^{2}\rangle\)
      & \(\gamma+\gamma\) & \(\mathrm{T}\)
      & \textbf{Ref.} \\
      \midrule
      Static                & \cmark & \xmark & \xmark & \xmark
      &\xmark &\xmark  & \cite{heisenberg1929folgerungen, Schwinger1951}  \\
      Inhomogeneous          & \cmark & \xmark & \cmark &\xmark &\xmark & \xmark & \cite{Hebenstreit:2011pm} \\
      Dynamically assisted   & \cmark & \cmark & \xmark & \xmark &\xmark & \xmark & \cite{Brezin:1970xf, Kim:2007pm}\\
      Thermally assisted     & \cmark & \xmark & \xmark & \xmark &\xmark & \cmark & \cite{Hallin:1994ad} \\
      Breit--Wheeler         & \xmark & \xmark & \xmark &\xmark & \cmark &\xmark &\cite{Breit:1934zz} \\
      \midrule
      \addlinespace[3pt]
      \textbf{Stochastic}   & \xmark & \xmark & \xmark & \cmark & \xmark & \xmark & \\
      \bottomrule
    \end{tabular}
  \end{threeparttable}
  \caption{Summary of Schwinger-type pair-creation mechanisms, indicating the presence (\cmark) or absence (\xmark) of specific background field configurations. Deterministic (grey) and stochastic (light blue) fields are distinguished by coloured shading. This classification separates classical backgrounds from purely quantum processes driven by real quanta. It is important to note that the stochastic channel presented in this work is distinct from all others. Like the Breit–Wheeler process, it is accessible perturbatively, but it is driven not by real particles or thermal baths, but by classical stochastic backgrounds.}
  \label{tab:processes}
\end{table}

In this work, we put forward the concept of the Schwinger effect in a stochastic gauge-field background. 
Such configurations can naturally arise in a range of high-energy phenomena, 
acting as transient stochastic sources that trigger vacuum decay into pairs of charged particles. 
Astrophysical environments and the early Universe naturally provide test-beds for such processes, where stochastic gauge-field fluctuations are expected to emerge. Yet, static field configurations of magnitudes comparable to those required for the static Schwinger mechanism are unlikely to occur in realistic settings. As a concrete example, axion–inflation was first identified in~\cite{Lozanov:2018kpk} as a natural setting for the static Schwinger effect in the early Universe. Subsequent works~\cite{Mirzagholi:2019jeb, Maleknejad:2019hdr, Domcke:2021fee, vonEckardstein:2024tix, Iarygina:2025ncl} further developed and extended this framework, exploring various aspects of the static Schwinger effect in axion–gauge systems. However, lattice simulations of reheating~\cite{Adshead:2015pva,Adshead:2023mvt} and inflation~\cite{Caravano:2021bfn,Figueroa:2023oxc,Sharma:2024nfu} reveal that the gauge fields generated during these epochs become highly stochastic. It is therefore crucial to formulate a stochastic description capable of capturing the intrinsic dynamics of pair production in realistic settings.

We develop a framework for pair production in stochastic Abelian gauge-field backgrounds, accounting for both stationary and non-stationary stochastic sources. To distill the essential physics, we focus on flat spacetime and zero temperature, where the process is non-thermal and the spacetime curvature is negligible compared to the characteristic dynamical scales. We then illustrate how the mechanism operates through three phenomenological examples within both standard model (SM) and beyond (BSM). The mechanism discussed here can be viewed as part of a broader class of stochastic pair-production phenomena, in which random field fluctuations provide the necessary conditions for vacuum decay into charged pairs. 
Notable examples include pair creation during cold inflation driven by the chiral anomaly~\cite{Maleknejad:2020pec, Maleknejad:2020yys}, 
the gravitational ABJ anomaly~\cite{Alexander:2004us, Maleknejad:2014wsa, Maleknejad:2016dci, Maleknejad:2024vvf}, 
and  the recently unveiled stochastic particle production sourced by cosmic perturbations ~\cite{Maleknejad:2024ybn, Maleknejad:2024hoz, Garani:2024isu, Garani:2025qnm, Eroncel:2025qlk}.

The structure of this paper is as follows. We begin in \cref{Sec:Particle_creation_and_QFT} with a review of the static Schwinger mechanism. This sets the foundation for \cref{Sec:Stochastic_Scwhinger_Effect}, where we introduce our central result: a stochastic extension of the Schwinger effect. In this section, we compute the effective action and pair production rate in the presence of a randomly fluctuating background of Abelian gauge fields. Building on this, sections \ref{Sec:Phenomenology_I} and \ref{Sec:Phenomenology_II} explores the phenomenological implications of the mechanism, focusing on its relevance for high-energy astrophysical environments and early-Universe cosmology. In \cref{sec:static-vs-stochastic} we compare the efficiency of vacuum decay in the static and stochastic Schwinger effects. Finally, \cref{Sec:conclusion} provides a summary of our findings and outlines directions for future investigation. Additional technical details and conventions are provided in the appendices.


\section{Setup: Charged Matter in Abelian Gauge Theories\label{Sec:Particle_creation_and_QFT}}
We start by setting up the framework and we adopt the mostly positive metric signature.  Specifically, we consider classical actions in flat spacetime for matter fields charged under a $U(1)$ gauge field $A_\mu$, i.e. 
\begin{align}
   S[X,X^*,A_\mu] = \int d^4x \, (\mathcal{L}_{G}+\mathcal{L}_{M}), 
\label{Eq:sQED_Lagrangian}
\end{align}
where $X$ is a matter field with $X^*$ denoting its conjugate representation, $\mathcal{L}_G$ contains the gauge sector and $\mathcal{L}_M$ the matter sector. Here, $\mathcal{L}_G$ can refer to Maxwell theory
\begin{align}
   \mathcal{L}_{A} = -\tfrac{1}{4}F_{\mu\nu}F^{\mu\nu} - \tfrac{1}{2\xi} (\partial_\mu A^\mu)^2, 
\label{Eq:Maxwell_Lagrangian}
\end{align}
where $F_{\mu\nu}$ is the field-strength tensor and the second term implements covariant gauge fixing \cite{Schwartz_2013}. In what follows, we employ Feynman gauge ($\xi\to 1$). Alternatively, $\mathcal{L}_G$ may describe a massive vector field as
\begin{align}
   \mathcal{L}_{A'} =- \frac{1}{4}F'_{\mu\nu}F'^{\mu\nu} - m_{A'}^2 A'_{\mu} A'^{\mu}, 
\label{Eq:Proca_Lagrangian}
\end{align}
where $A'$ can be understood as a heavy photon with mass $m_{A'}$. This framework is of particular importance in cosmology and astrophysics, where dark photons can serve as a dark matter background coupled to charged particles \cite{Fabbrichesi_2021, Caputo:2021eaa}. 
The matter Lagrangian $\mathcal{L}_{M}$ can be either a complex scalar or a charged fermion as
\begin{align}
   \mathcal{L}_{\phi} &= - (D_\mu \phi)(D^\mu \phi)^* - m^2|\phi|^2, 
\label{Eq:complexScalar_Lagrangian}\\
   \mathcal{L}_{\psi} &= i\bar\psi(\slashed{D}-m)\psi,
\label{Eq:Dirac_Lagrangian}
\end{align}
with the gauge-covariant derivative $D_\mu=\partial_\mu+ig Q A_\mu$ and $\slashed{D}=\gamma^{\mu}D_{\mu}$, where $\gamma^{\mu}$ are the Dirac gamma matrices. Here $g$ is the gauge coupling and $Q$ is the charge of the matter field.  Throughout, the scalar field is taken as a millicharged BSM species, while the spinor may be a SM fermion or another millicharged species. Such millicharged particles can naturally give rise to viable dark matter candidates~\cite{Bogorad:2021uew,Maleknejad:2022gyf}. The Abelian stochastic background is either the electromagnetic (EM) field or a dark photon.

To study the quantum dynamics of our theory, we define a partition function by coupling the fields to external sources and integrating over all field configurations,  
\begin{align}
    \mathcal{Z}[\{J\}] = \int \mathcal{D}X\,\mathcal{D}X^*\,\mathcal{D}A_\mu\ 
    e^{i \int d^4x\ \big( \mathcal{L} + X J_X^* + X^* J_X + A_\mu J^\mu \big) },
\label{Eq:FullPartitionFunction}
\end{align}
where $\{J\}$ the set of sources. The corresponding generating functional for connected diagrams is  
$    \mathcal{W}[\{J\}] = -i \ln \mathcal{Z}[\{J\}]$. Making use of the functional Legendre transform, we find the quantum effective action to be \cite{JonaLasinio1964}
\begin{equation}
    \Gamma[\bar X,\bar X^*,\bar A_\mu]=\mathcal{W}[\{J\}]-\int d^4x\ J_I(x)\ \frac{\delta \mathcal{W}[\{J\}]}{\delta J_I(x)},
\label{Eq:EffectiveActionFunctional}
\end{equation}
where $\bar X=\langle \hat X\rangle$ denotes the expectation value of field $X$, the index $I$ labels each element in set ${\{J\}}$ and there is an implicit Einstein summation over them.

\subsection{Vacuum Instability and Pair Production \label{Sec:Pair_Production}}

In this section, we outline the mathematical framework for particle production in interacting quantum field theories. Such phenomena arise from violations of global energy conservation due to couplings to dynamical background fields, which break the full Poincaré invariance of Minkowski spacetime. For spatially homogeneous and isotropic sources, the symmetry is reduced to the six generators of spatial translations and rotations. The residual spatial translations then constrain the produced states to appear in pairs of equal and opposite momenta, $\bq$ and $-\bq$, and $U(1)$ charge conservation enforces opposite charges.

Even when these global symmetries are broken, for example by spatially varying sources, local Lorentz invariance ensures that energy-momentum conservation holds in each local inertial frame. Particle creation at a point thus respects the usual kinematic constraints locally, while on larger scales the background can exchange momentum and energy with the produced pairs. This exchange may lead to redshifting or mode mixing, distorting the spectrum expected in the homogeneous case.

In what follows, we present the effective action approach, which will serve as our primary framework to capture such vacuum instabilities via quantum corrections. In QFT, the vacuum persistence amplitude measures vacuum stability under external fields and quantum effects. It is given by
\begin{align}
    \mathcal{A} = \bra{0_\text{in}} e^{-i t \hat{H}} \ket{0_\text{in}} = \braket{0_\text{out} | 0_\text{in}},
\end{align}
where $\ket{0_{\text{in}}}$ denotes the asymptotic \emph{in} vacuum of the quantum matter field, $t$ is the physical time and $\hat{H}$ is the system's Hamiltonian. In the absence of external currents ($J=0$), the vacuum is time-translation invariant and the overlap satisfies $\mathcal{A}=1$. When sources are present ($J\neq0$), this symmetry is broken and the \emph{in} and \emph{out} vacua may differ.\footnote{Note that a non-zero $J$ is not a sufficient condition for vacuum decay. Particle production occurs if and only if the effective action develops an imaginary part.}  The vacuum persistence probability is
\begin{align}
    \mathcal{P} &= \big|\braket{0_\text{out}|0_\text{in}}\big|^2
    = e^{-2\,\mathrm{Im}\Gamma}
    = \exp\!\left[-2\int d^4x\, w(x)\right],
\end{align}
where $\Gamma$ is the effective action in the background field and $w(x)$ can be interpreted as a local pair-creation rate density. A non-zero $\mathrm{Im}\,\Gamma$ thus signals vacuum instability. The complementary probability of decay is  
\begin{align}
    \mathcal{P}_\text{decay} = 1 - \mathcal{P} \;\simeq\; 2\,\mathrm{Im}\,\Gamma
    = 2 \int d^4x\, w(x),
\end{align}
in agreement with the optical theorem, which relates forward scattering to total production rates \cite{Schwinger1951, ItzyksonZuber1980, Peskin:1995ev}. In the perturbative regime, where all mode occupations are small, the vacuum persistence probability is $\mathcal{P} \;\approx\; \exp\!\left[-\int d^3\bq\, N_{\bq}\right]$,
with $N_{\bq}=\bra{0_{\rm in}}\,\hat a_{\bq}^\dagger \hat a_{\bq}\,\ket{0_{\rm in}}$ the particle number in mode $\bq$.
This corresponds to the total number of produced pairs, so that the pair number density is
\begin{align}
n_{\rm pairs}\simeq \frac{1}{V}\,\mathcal{P}_{\rm decay}, 
\qquad 
n_{\rm particles}\simeq 2\,n_{\rm pairs}.
\label{eq:n-pairs-}
\end{align}
This relation holds only in the weak-field limit; in strong fields, $\mathcal{P}_{\rm vac}$ alone does not determine the full spectrum $N_{\bq}$.

The effective action of scalar or fermion QED in a background field is obtained from the connected generating functional \cref{Eq:FullPartitionFunction} by treating the gauge field $A_\mu$ as classical, setting the external sources $\{J\}$ to zero, and integrating out the matter fields. At one-loop order, the action naturally separates as \cite{Schwinger1951,ItzyksonZuber1980}
\begin{align}
    \Gamma[A_\mu] 
    = S_{\text{YM}}[A_\mu] + \Gamma_{\text{1-loop}}[A_\mu], 
\label{Eq:EffectiveAction_Decomposition}
\end{align}
where $S_{\text{YM}}[A_\mu]$ is the classical Maxwell action (with gauge fixing), $\Gamma_{\text{1-loop}}[A_\mu]$ encodes the quantum corrections from integrating out matter fields, and we have omitted the bar in $\bar A_{\mu}$ for notational simplicity. For scalar and fermionic QED, one finds  
\begin{align}
    \Gamma_{\text{1-loop}}^{b}[A_\mu] 
    &= i \ln \det\!\left(\frac{D_\mu D^\mu - m^2}{\Box - m^2}\right),
\label{Eq:OneLoopBoson}
\\[6pt]
    \Gamma_{\text{1-loop}}^{f}[A_\mu] 
    &= -{{i}} \ln \det\!\left(\frac{i\slashed{D} - m}{i\slashed{\partial} - m}\right),
\label{Eq:OneLoopFermion}
\end{align}
where the D'Alembertian is $\Box \equiv \partial^\mu \partial_\mu$. The bosonic functional determinant arises from integrating out charged scalars, while the fermionic one corresponds to integrating out Dirac fields and includes an additional trace over spinor indices. A convenient tool to evaluate these is the proper-time representation (see \cref{app:eq-In-Det}). For the scalar case, one has  
\begin{align}
    \Gamma_{\text{1-loop}}^{b}[A_\mu] 
    &= -\int_0^\infty \frac{ds}{s}\, e^{-is(m^2 - i\varepsilon)} 
    \int d^4x \,\Big[\langle x \,|\, e^{is D^2} \,|\, x \rangle 
    - \langle x \,|\, e^{is\Box} \,|\, x \rangle \Big].
\label{Eq:ProperTime_Scalar}
\end{align}
The fermionic one-loop effective action is obtained by replacing 
$D^2$ with the squared Dirac operator $\slashed{D}^2 = D^2 + \tfrac{gQ}{2} F_{\mu\nu}\,\sigma^{\mu\nu}$ and multiplying the effective action by a factor of $\tfrac{1}{2}$.  Here $\sigma^{\mu\nu} \equiv \tfrac{i}{2}[\gamma^\mu,\gamma^\nu]$ 
denotes the antisymmetric generators of the Lorentz group in the spinor representation.

\subsection{Static Schwinger Effect and Euler--Heisenberg Theory\label{Sec:Euler_Heisenberg}}

For photon energies $\omega \ll m$, quantum corrections to Maxwell theory are described by the Euler--Heisenberg (EH) effective action, obtained by integrating out charged particles in constant electromagnetic fields \cite{heisenberg1929folgerungen}. The theory admits an expansion in the invariants $F_{\mu\nu}/m^2$ and $\tilde F_{\mu\nu}/m^2$, capturing non-linear QED effects such as the (static) Schwinger effect.

For a vanishing magnetic field, $B=0$, and constant electric  field $E$, the bosonic and fermionic imaginary parts of the EH density are \cite{heisenberg1929folgerungen, Schwinger1951}
\begin{align}
2\Im\mathcal{L}_b
&= \frac{g^2Q^2E^2}{8\pi^3}\sum_{n=1}^\infty \frac{(-1)^{n+1}}{n^2}
\exp\!\left[-\frac{n\pi m^2}{gQE}\right],
\label{eq:Schwinger-scalar} \\[6pt]
2\Im\mathcal{L}_f
&= \frac{g^2Q^2E^2}{4\pi^3}\sum_{n=1}^\infty \frac{1}{n^2}
\exp\!\left[-\frac{n\pi m^2}{gQ E}\right].
\label{eq:Schwinger-fermion}
\end{align}
When both invariants are non-zero, one may choose $\mathbf{E}\parallel \mathbf{B}$, yielding \cite{heisenberg1929folgerungen, Schwinger1951}
\begin{align}
2\,\mathrm{Im}\,\mathcal{L}_b
&= \frac{g^2Q^2EB}{2(2\pi)^2}
\sum_{n=1}^\infty \frac{(-1)^{n+1}}{n}\,
\text{csch}\!\left(\frac{n\pi B}{E}\right)\,
\exp\!\left[-\frac{n\pi m^2}{gQE}\right],
\label{eq:EH-scalar}\\  
2\,\mathrm{Im}\,\mathcal{L}_f
&= \frac{g^2Q^2EB}{(2\pi)^2}
\sum_{n=1}^\infty \frac{1}{n}\,
\coth\!\left(\frac{n\pi B}{E}\right)\,
\exp\!\left[-\frac{n\pi m^2}{gQE}\right].
\label{eq:EH-fermion} 
\end{align}
These series, which encode the rate of particle creation per unit four-volume, may also be interpreted as a sum over instanton  contributions~\cite{Kim:2003qp}. The  validity of the Euler--Heisenberg effective action is restricted to static EM backgrounds or the regime $\omega \ll m$. Extensions to deterministic backgrounds with non-vanishing expectation value, $\langle F_{\mu\nu}\rangle\neq 0$, such as laser fields, have also been investigated. In these cases, the effective action must be generalised to account for spatial or temporal inhomogeneities of the external field. Such generalisations provide a framework for studying more realistic situations, including localised or pulsed field configurations~\cite{Brezin:1970xf,Narozhny:1970uv,Dunne:2004nc}.


\section{Stochastic Schwinger Effect\label{Sec:Stochastic_Scwhinger_Effect}}

In this section we develop a general framework for extending the standard Schwinger formalism to vacuum decay in the presence of stochastic fields. Unlike the deterministic classical backgrounds discussed in Section~\ref{Sec:Particle_creation_and_QFT}, stochastic fields are characterised by a vanishing expectation value but non-trivial correlations, thereby capturing the intrinsically random fluctuations relevant in astrophysical and cosmological settings. These fluctuations are described statistically, through spectral densities and their initial conditions~\cite{Weinberg:2008zzc}. In this case, the differential operator appearing in the effective action is no longer linear, which prevents a straightforward non-perturbative solution. Instead, we perform an expansion in the coupling $g$, expressing the effective action as a series in powers of the field. This computation, based on the Schwinger proper-time method, remains non-perturbative in the mass $m$ but perturbative in the coupling $g$ (or $e$ in QED). Thus, its nature is distinct from the Schwinger–DeWitt expansion \cite{Schwinger1951, DeWitt:1964mxt}, which is perturbative in $\Box/m^2$.

To treat stochastic background fields systematically, we promote $\hat{A}_\mu$ to an operator-valued stochastic process with Gaussian statistics,  
\begin{equation}
    \langle \hat{A}_\mu(x)\rangle_s = 0, 
    \qquad 
    \langle \hat{A}_\mu(x) \hat{A}_\nu(y)\rangle_s = G_{\mu\nu}(x-y),
    \label{Eq:Gaussianity_SSE}
\end{equation}
where $\langle \dots \rangle_s$ denotes the stochastic average over background field realisations. A real vector field may then be expanded in momentum space as  
\begin{equation}
    \hat{A}_{\mu}(x)=\int d^4 q\ \hat{A}_\mu(q)\, e^{iq\cdot x} ,
\label{Eq:FieldDecomposition_SSE}
\end{equation}
with mode decomposition  
\begin{equation}
    \hat{A}_{\mu}(q)=\sum_{\sigma} e^\sigma_\mu(\hat \bq) 
    \left[\Theta(q^0)\,\delta(q^0 - \omega_{\bq,\sigma})\,\alpha_{\bq,\sigma}\, A_{\bq,\sigma} 
   + \Theta(-q^0)\,\delta(q^0 + \omega_{\bq,\sigma})\,\alpha_{-\bq,\sigma}^{*}\,A_{-\bq,\sigma}^{*}\right],
\label{EQ:FieldOperator_SSE}
\end{equation}
where $\omega_{\bq,\sigma}$ is the dispersion relation of the mode at polarisation state $\sigma$ and wave vector $\bq$, $e^\sigma_\mu(\hat \bq)$ are the physical polarisation vectors, and the Heaviside step function $\Theta(x)$ ensures propagation along future-pointing trajectories. For modes propagating in the $\hat{\bq}=-\hat{\boldsymbol{r}}$ direction, the three polarisation states take the form  
\begin{align}
e^{\pm}_{\mu}(\hat{\bq}) = \tfrac{1}{\sqrt{2}}(\hat{\theta} \pm i \hat{\phi}), 
\qquad  
e^{0}_{\mu}(\hat{\bq}) = \hat{\bq},  
\label{eq:hat-A}
\end{align}
with $\pm$ and $0$ denoting transverse and longitudinal polarisations, respectively.\footnote{While free, on-shell photons in vacuum propagate only two transverse polarisations, the presence of a mass term (as in a Proca field) or a thermal or plasma background generally gives rise to a non-zero longitudinal mode.} The mode amplitudes $A_{\bq,\sigma}$ are complex coefficients, while $\{\alpha_{\bq,\sigma}\}$ are independent random variables specifying the stochasticity of each mode. Their statistics are defined by  
\begin{equation}
    \langle \alpha_{\bq,\sigma}\, \alpha^*_{\bq',\sigma'}\rangle_s 
    = \delta_{\sigma\sigma'}\,\delta^3(\bq-\bq').
    \label{Eq:Commutator_SSE}
\end{equation}

We now turn to the QED effective action. Here we start with the complex scalar field, $\phi$, which we treat quantum mechanically and integrate out in the presence of a stochastic vector background. In analogy with the deterministic case, the effective action is obtained from the logarithm of the functional determinant introduced in \cref{Eq:OneLoopBoson}, but with the crucial distinction that the background is now a stochastic process rather than a fixed configuration. All physical quantities must therefore be averaged both over the quantum vacuum of the matter field and over the stochastic  ensemble of the background. To streamline notation, we define
\begin{align}
\llangle \dots \rrangle \equiv 
\langle \bra{0_{\text{in}}}\,\dots\,\ket{0_{\text{in}}}\rangle_s
= \int \mathcal{D}A\, P[A]\,
\bra{0_{\text{in}}}\,\dots\,\ket{0_{\text{in}}}_A ,
\end{align}
where the subscript $A$ indicates evaluation in a given background realisation, and $P[A]$ is the probability functional characterizing the stochastic ensemble. This encodes the interplay between quantum fluctuations of matter and random fluctuations of the background, and sets the stage for evaluating the trace in the effective action. The derivation follows the general formalism in \cite{ItzyksonZuber1980}, adapted to the stochastic gauge-field background considered in this work. Here, we consider again a quantised complex scalar field in the presence of a background EM field. 
\begin{align}
    \Gamma_{\text{1-loop}}^{b}[A_\mu] &=i \text{Tr}\ln\left[\frac{(\hat{p}-g Q \hat{A}({x}))^2+m^2 - i \epsilon}{\hat{p}^2+m^2 - i \epsilon}\right].
\end{align}
Above, $\Tr$ denotes the functional trace, which includes sums over both discrete indices (e.g., spinor, gauge, or other internal indices) as well as integrals over continuous spacetime or momentum variables. For simplicity, we now define the Hermitian operators,  $ f(A_\mu) $ and $ \hat{O}(A_\mu)$, as follows 
\begin{align}
    \hat f(A) &=  \hat{p}_{\mu} \hat{A}^{\mu} + 2  \hat{A}^{\mu} \hat{p}_{\mu} - g Q \hat{A}_{\mu}\hat{A}^{\mu}, \\
    \hat {O}(A) &= gQ \hat f(A)+gQ \hat f(A)\frac{1}{\hat{p}^2+m^2-i\epsilon}\hat{O}(A),
    \label{eq:mathcal-O}
\end{align}
where $\hat{p}_{\mu}=i\p_{\mu}$ is the momentum operator. Next, we define the positive (negative) energy mass-shell delta function as
\begin{equation}
  \hat{\delta}^{\pm}(\hat{p}) =  \delta(\hat{p}^2+m^2) \Theta(\pm \hat{p}^0).
  \label{eq:pm-delta}
\end{equation}
The effective action takes the following form, where details of its derivation are provided in Appendix \ref{App:Calculations},
\begin{equation}
 \text{Im}\Gamma_{\text{1-loop}}^{b}[A_\mu]  = -\text{Tr}\ln\left(  \hat{\mathbb{I}}  - \hat{O}(A) \, \hat{\delta}^+ \hat{O}(A) \, \hat{\delta}^- \right).
 \label{sec:QED-expansion}  
\end{equation}
Expanding the above expression perturbatively to order $g^2$, we obtain the decay probability to the charged scalar as
\begin{equation}
    \mathcal{P}^b_{\text{decay}} =-g^2Q^2\Tr[(2\hat{A}_\mu\partial^\mu+ \partial^\mu \hat{A}_\mu)\hat{\delta}^+(2\hat{A}_\nu\partial^\nu+\partial^\nu \hat{A}_\nu)\hat{\delta}^-] + \mathcal{O}(g^3).    
\label{eq:P-decay}
\end{equation}
For the fermionic case, the vacuum decay probability can be written in a similar form \cite{Schwinger1951}
\begin{equation}
    \mathcal{P}^{f}_{\text{decay}}
    = -\, g^2Q^2\, \Tr\!\left[\slashed{A}\,\delta^+\,\slashed{A}\,\delta^-\right] 
    + \mathcal{O}(g^3).
    \label{eq:P-decay-fermion}
\end{equation}

\subsection{Stationary background}

When the background gauge field arises from a stationary process, such as a source with time-translation invariance, it admits a well-defined Fourier expansion in terms of four-momentum. In this case, the modes are stationary, each characterised by a definite frequency, as a result
\begin{equation}
\bra{p} \hat{A}_{\mu}({x}) \ket{p-q} =  \hat{A}_{\mu}(q)
= \frac{1}{(2\pi)^4} \int d^4x \, \hat{A}_{\mu}(x)\, e^{-iq\cdot x}.
\end{equation} 
By inserting \cref{eq:rho-pm} into \cref{eq:P-decay}, the decay probability appearing in \cref{eq:P-decay} takes the form
\begin{align}
\begin{split}
        \mathcal{P}^b_{\text{decay}} 
        &= \frac{g^2Q^2}{(2\pi)^4} 
        \int_q \,  \llangle \hat{A}_{\mu}(q)\, \hat{A}_{\nu}(-q)\rrangle 
        \int\!\!\!\int_{p_1,p_2} \delta^{(4)}(q-p_1+p_2)\delta(p_1^2+m^2) \delta(p_2^2+m^2) \\
        &
        \quad\times \Theta(p_1^0) \Theta(-p_2^0) \,(2p_{1}-q)^{\nu} (2p_2+q)^{\mu},
        \label{eq:W1-int}
\end{split}
\end{align}
where, for shorthand, we denote $\int_q \equiv \int d^4q$. Here we summarise only the final expressions for the decay into scalar particles; the detailed derivation leading to this result is presented in Appendix \ref{App:Calculations}. Using \cref{eq:app-momentum,eq:app-ff}, we obtain the one-loop vacuum decay probability into charged bosons induced by a stationary, stochastic gauge field as
\begin{align}
  \mathcal{P}_{\text{decay}}^b
  = \frac{g^2Q^2\pi}{3} \int_q 
  \Theta\!\left(- q^2 - 4m^2\right)
  \left(1 + \frac{4m^2}{q^2}\right)^{\!3/2} 
  \llangle -\hat{F}_{\mu\nu}(q)\, \hat{F}^{\mu\nu}(-q) \rrangle.
  \label{eq:Gamma-Boson}
\end{align}
This perturbative contribution vanishes for a strictly constant electromagnetic field, as expected. The spectral number density of produced scalar particles can be read directly from \cref{eq:n-pairs-} as
\begin{align}
\begin{split}
\llangle n^b({\omega}) \rrangle
&= \frac{g^2Q^2\pi}{3V} 
\int d^3\bq \ \Theta\!\left(\omega^2 - |\bq|^2 - 4m^2\right)\,\left(1 - \frac{4m^2}{\omega^2 - |\bq|^2}\right)^{3/2}\\ &\quad\times
\llangle -F_{\mu\nu}(\omega,\bq)\, F^{\mu\nu}(-\omega,-\bq) \rrangle,
  \label{eq:n-omega}
\end{split}
\end{align}
where $q^\mu = (\omega, \bq)$. 
Using \cref{eq:n-omega-fermion}, the corresponding spectral number density for fermionic pair production takes the form
\begin{align}
\begin{split}
\llangle n^f({\omega}) \rrangle
&= \frac{4g^2 Q^2\pi}{3V} 
\int d^3\bq\ \Theta\!\left(\omega^2 - |\bq|^2 - 4m^2\right)\, 
\left(1 + \frac{2m^2}{\omega^2 - |\bq|^2}\right)\, \left(1 - \frac{4m^2}{\omega^2 - |\bq|^2}\right)^{1/2}\\
& \quad\times \llangle -F_{\mu\nu}(\omega,\bq)\, F^{\mu\nu}(-\omega,-\bq) \rrangle.
 \label{eq:n-omega-fermion-}
\end{split}
\end{align}
Both distributions feature a clear kinematic threshold, $\omega^2 > |\bq|^2 + 4m^2$, indicating that only background modes with energy above twice the particle's rest mass contribute to pair creation. 
The number densities scale as $g^2 Q^2$, consistent with the perturbative coupling between the charged fields and the background gauge fluctuations. 
Furthermore, the production rate is directly proportional to  the spectral intensity of the underlying field fluctuations.

\subsection{Non-stationary background} \label{sec:non-stationary}

In this section, we develop a framework suited to transient and non-stationary backgrounds, 
such as those arising during preheating, magnetogenesis, or phase transitions in the early Universe, 
and in astrophysical environments like magnetars, pulsar winds, and turbulent plasmas.  When the background gauge field originates from a time-dependent process, the modes no longer possess well-defined frequencies. As a result, non-stationary fields can be Fourier-decomposed in terms of three-momentum \cite{Weinberg:2008zzc} as
 \begin{align}
   \hat{A}_{\mu}(x) = \sum_{\sigma} \, \int d^3\bq\,   \left[ \, A_{\mu,\sigma}(t,\bq) \hat{\alpha}_{\bq,\sigma} + h.c. \right] \, e^{i\bq\cdot\bx}.
\end{align}
Alternatively, a natural extension of the four-momentum Fourier transform for such fields is the windowed Fourier transform, commonly referred to as the short-time Fourier transform (STFT). It is defined as a Fourier transform localised in both time and frequency domains. Mathematically, the STFT of a mode function $ A_{\mu}(x)$ is given by \cite{Cohen1995, Flandrin1999}
\begin{align}
    \hat{A}_{\mu}(q;t) = \frac{1}{(2\pi)^4}\int d^4x \, \hat{A}_{\mu}(x) \, W(x^0 - t) \, e^{-i q\cdot x},
   \end{align}
where $W(x^0 - t)$ is a timelike window function centred at time $t$. The STFT of a gauge field can be expressed as
\begin{equation}
    \hat A_{\mu}(q;t)=\sum_{\sigma} e^\sigma_\mu(\hat \bq) \left[\Theta(q^0)  \hat{\alpha}_{\bq,\sigma} A_{\sigma}(q;t)  +\Theta(-q^0)  \hat{\alpha}_{-\bq,\sigma}^{*} A_{\sigma}^{*}(-q;t)\right],
    \label{eq:A-non-stationary}
\end{equation}
where the mode function $A_{\sigma}(q;t)$ is defined in terms of the Fourier transform as
\begin{align}
A_{\sigma}(q;t) = \frac{1}{2\pi} \int_{-\infty}^{+\infty} dx^0 \, A_{\sigma}(\bq,x^0) \, W(x^0 - t) \, e^{i q^0x^0}.
\label{eq:A-non-stationary-mode}
\end{align}

The window function localises the mode around a given time $t$, and the corresponding Fourier transform yields a local estimate of its instantaneous frequency content. This should be contrasted with stationary backgrounds, where the Fourier transform is globally defined over the entire time domain. In the stationary case the mode function is $A_{\sigma}(\bq,x^0)=A_{\bq,\sigma} e^{-i \omega_{\bq} x^0}$, and the window function is not needed, i.e., $W(t'-t)=1$.  In this case, \cref{eq:A-non-stationary} reduces to \cref{EQ:FieldOperator_SSE}. 
By contrast, in the non-stationary case, the window function broadens this delta distribution, smearing the frequency support and reflecting the intrinsic time dependence of the background.

\begin{figure}
    \centering
\includegraphics[width=\linewidth]{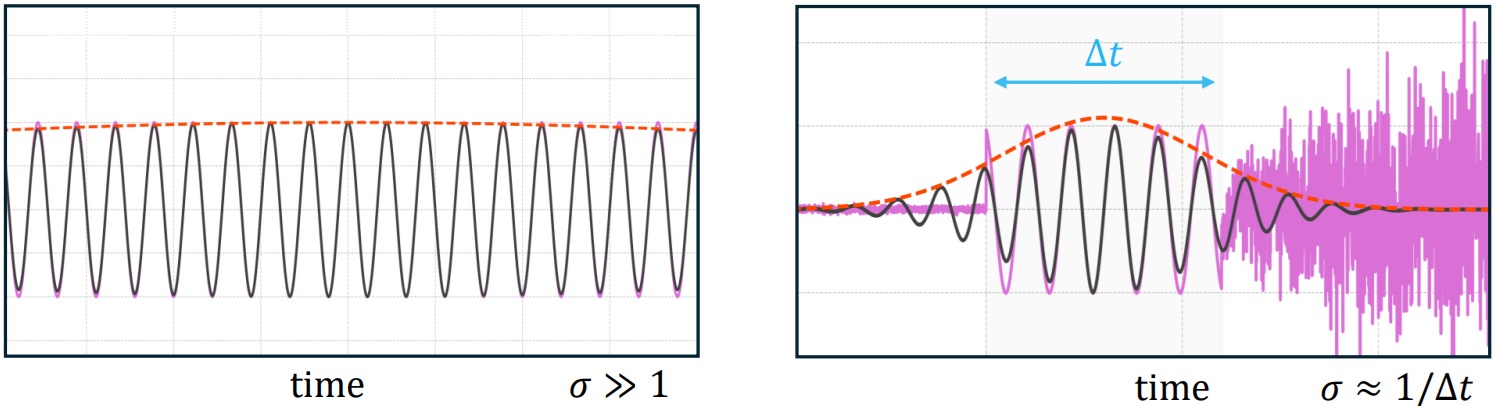}
  \caption{Illustration of the short-time Fourier transform (STFT) with Gaussian windows. 
The orchid curve is the signal, the dashed orange curve the Gaussian window function, and the solid black curve the resulting windowed signal.
The left panel shows the case $\sigma \gg 1$, where the wide Gaussian window leaves the sinusoidal signal essentially unchanged, corresponding to the global Fourier transform. 
The right panel shows a finite-width Gaussian window ($\sigma \approx 1/\Delta t$), which localises the analysis in time and demonstrates how the STFT resolves transient features. }
    \label{fig:placeholder}
\end{figure}

Among the commonly employed window functions in short-time Fourier analysis are the rectangular window, which provides sharp time localisation at the expense of frequency leakage, and the Gaussian window, which is often preferred in physics due to its optimal balance between time and frequency resolution as dictated by the uncertainty principle. In this work, we employ a Gaussian window function,
\begin{align}
    W(t-t') =   e^{-\frac{(t-t')^2}{2\sigma^2} },
\end{align}
where $\sigma$ determines the temporal width of the Gaussian window. 
For a system whose analytical solution is valid only within a short-time interval $\Delta t$, 
it is natural to set $\sigma \approx \Delta t$, ensuring that the window captures the relevant dynamics without truncation or overextension. 
This choice embodies the time--frequency uncertainty relation, 
$\Delta t \approx \sigma$ and $\Delta \omega \approx 1/\sigma$~\cite{Cohen1995,Feichtinger1998, Flandrin1999, Oppenheim2009} ; see Fig. \ref{fig:placeholder}.

In this part, we focus on small $\sigma$, corresponding to the short-time resolution required to capture transient non-stationary dynamics and rapidly varying field configurations, at the cost of reduced frequency resolution. In this regime, the STFT reduces to the Gabor transform, which optimally balances time and frequency resolution~\cite{Cohen1995,Feichtinger1998,Oppenheim2009}. One may further generalise to a momentum-dependent width, $\sigma=\sigma_{\bq}$, matched to the oscillation timescale of each mode, $\sigma_{\bq}\sim\omega_{\bq}^{-1}$, in analogy with adaptive wavelet methods~\cite{Mallat1999,Daubechies1992}. Within this framework, the equal-time two-point function of the field strength, built from a pair of Gabor transforms, is equivalent to a Gaussian-smeared Wigner transform of the underlying correlator. 

For non–stationary backgrounds the spectral number density is controlled by the unequal–time
field–strength correlator. From \cref{eq:n-pairs-} we obtain
\begin{align}
\begin{split}
\llangle n^b({\omega}) \rrangle
&= \frac{2 g^2 Q^2 \pi}{3 (2\pi)^2 V} 
\int d^3\bq \ \Theta\left(\omega^2-|\bq|^2-4m^2\right) 
\left(1 - \frac{4m^2}{\omega^2-|\bq|^2}\right)^{3/2}
\\
&\quad \times \int\! dt' \int\! dt'' \,
e^{-\frac{(t'-t)^2+(t''-t)^2}{2\sigma^2}}\,
e^{i q^0 (t'-t'')} \,
\llangle -\hat F_{\mu\nu}(t'',\bq)\, \hat F^{\mu\nu}(t',-\bq) \rrangle,
\label{eq:n-omega-non-stationary}    
\end{split}
\end{align}
with 
$\llangle \hat F_{\mu\nu}(t'',\bq)\, \hat F^{\mu\nu}(t',-\bq) \rrangle
= (q^2 \eta^{\mu\nu} - q^\mu q^\nu)
\llangle \hat A_{\mu}(t'',\bq)\, \hat A_{\nu}(t',-\bq) \rrangle$.
The fermionic case follows from \cref{eq:n-omega-fermion} as
\begin{align}
\llangle n^f({\omega}) \rrangle
&= \frac{8 g^2 Q^2 \pi}{3 (2\pi)^2 V}
\int d^3\bq \ \Theta\left(\omega^2-|\bq|^2-4m^2\right) \,
\left(1 + \frac{2m^2}{\omega^2-|\bq|^2}\right)
\left(1 - \frac{4m^2}{\omega^2-|\bq|^2}\right)^{1/2}
\nonumber\\
&\quad \times \int\! dt' \int\! dt'' \,
e^{-\frac{(t'-t)^2+(t''-t)^2}{2\sigma^2}}\,
e^{i q^0 (t'-t'')} \,
\llangle -\hat F_{\mu\nu}(t'',\bq)\, \hat F^{\mu\nu}(t',-\bq) \rrangle.
\label{eq:n-omega-fermion-non-stationary}
\end{align}
Both spectra display the kinematic threshold $\omega^2>|\bq|^2+4m^2$, so only sufficiently energetic
background modes produce pairs. The number densities scale as $g^2 Q^2$, reflecting their perturbative
origin, and are proportional to the Gaussian–windowed unequal–time correlator of the gauge field,
which reduces to the stationary result when the correlator depends only on $t'-t''$. Notably, an analogous structure was recently identified in gravitational particle production in \cite{ Maleknejad:2024hoz} in which the cosmic perturbations background is intrinsically non-stationary, as cosmological expansion breaks time-translation invariance.

\section{Stochastic Schwinger in High Energy Astrophysics}\label{Sec:Phenomenology_I}

In this section we focus on two representative {stationary} stochastic backgrounds relevant to high–energy astrophysics: (i) electromagnetic modes in a cold medium and (ii) a dark–photon background. By stationary we mean approximate time–translation invariance, so the background admits a well–defined spectral decomposition. While we develop these two concrete cases, the stochastic Schwinger mechanism is more general and can equally be driven by non–stationary backgrounds, which we consider in the next section.

For later convenience, we note that electromagnetic field configurations are naturally classified using the two Lorentz invariants $I_1=-\frac12F_{\mu\nu} F^{\mu\nu}$ (scalar invariant) and $I_2=-\frac12F_{\mu\nu}\tilde{F}^{\mu\nu}$ (pseudoscalar invariant). The classification is summarised in Table \ref{tab:em_field_types}. Equation (\ref{eq:n-omega}) indicates that pair production is possible only when $I_1>0$, which can be either electric-like or mixed.

\begin{table}[H]
  \centering
  \small
  \begin{threeparttable}
    \setlength{\tabcolsep}{8pt}
    \renewcommand{\arraystretch}{1.3}
    \begin{tabular}{C C L}
      \toprule
      \(F_{\mu\nu}F^{\mu\nu}\) & \(F_{\mu\nu}\tilde{F}^{\mu\nu}\) & \textbf{Field type} \\
      \midrule
      \(>0\) & \(=0\)   & Electric-like \\
      \(<0\) & \(=0\)   & Magnetic-like \\
      \(=0\) & \(=0\)   & Null (radiation) \\
      any    & \(\neq 0\) & Mixed / general \\
      \bottomrule
    \end{tabular}
  \end{threeparttable}
  \caption{Classification of electromagnetic fields by the Loretnz invariants.}
  \label{tab:em_field_types}
\end{table}

In what follows, we consider  the constraint equation $\partial_{\mu} A^{\mu} = 0$, and remain agnostic regarding the underlying origin of the background, modelling the spectral energy density using a broken power law to allow sufficient flexibility to approximate realistic spectra. We decompose the electromagnetic field into transverse and longitudinal polarisations as
\begin{align}
 E_{\pm}(q) &= -i q^0 A_{\pm}(q),\\
 E_{L}(q) &= i  \frac{q^2 }{q^0} A_{L}(q)),\\
B_{\pm}(q) &= \mp i \lvert \bq \rvert A_{\pm}(q),
\end{align}
in which $ A_{L}(q)$ is the longitudinal mode defined as
\begin{align}
 A_{L}(q) \equiv A_{i}(q) \hat{q}^i. 
\end{align}
This decomposition leads to the gauge-invariant combination
\begin{align}
     -\hat{F}_{\mu\nu}(q) \hat{F}^{\mu\nu}(-q)  = -2q^2 \, \left( \sum_{\sigma=\pm} \hat{A}_{\sigma}(q) \hat{A}_{\sigma}(-q) - \frac{q^2}{(q^0)^2}  \hat{A}_{L}(q) \hat{A}_{L}(-q)\right).
     \label{eq:FF--}
\end{align}
We are interested in the regime where $I_1>0$, which requires either $-q^2>0$ (timelike photons) and/or the presence of a non-zero longitudinal mode. Using (\ref{eq:hat-A}) and (\ref{eq:int-AAA}), we obtain
\begin{align}
  \llangle  \hat{A}_{\sigma}(q) \hat{A}_{\sigma'}(-q) \rrangle = \frac{VT}{(2\pi)^4} \, \delta_{\sigma\sigma'} \, \delta(q^0 - \omega_{\bq,\sigma}) \,   P_{\sigma}(\omega_{\bp,\sigma}),
   \label{eq:AA-P}
\end{align}
where $ P_{\sigma}(\omega_{\bp,\sigma})= \lvert A_{\bq,\sigma}\rvert^2$  is the vector power spectrum. Explicitly, we write
\begin{align}
 P_{\sigma}(\omega)  = \frac{{\mathcal{C}}_{\sigma}}{\omega_\star} \begin{cases} 
  \big( \frac{\omega}{\omega_\star}\big)^{-2+\gamma_{\sigma}} & \Lambda_\text{IR} \leq \omega <\hspace{0.1cm} \omega_\star\\
   \big( \frac{\omega}{\omega_\star}\big)^{-\delta_\sigma} & \hspace{0.1cm}\omega_\star \hspace{0.1cm}\leq \omega \leq \Lambda_\text{UV} 
  \end{cases},
  \label{eq:spectrum--}
  \end{align} 
where $\mathcal{C}_\sigma$ is a dimensionless constant, and the exponents $\gamma_{\sigma}<\delta_{\sigma}$ are order one positive numbers. In realistic sources, the generated spectrum is truncated at low and high energies by inherent UV and IR frequency cutoffs set by the source environment, denoted as $\Lambda_\text{IR}$ and $\Lambda_\text{UV}$, respectively. These cutoffs are generally polarisation dependent.  Finally, the delta function in \cref{eq:AA-P} is used to specify the dispersion relation of each mode
\begin{align*}
   \omega^2_{\bq,\sigma} = \lvert \bq \rvert^2 +\Delta^2_{\sigma}(\bq). 
\end{align*}
Throughout this section, we neglect chiral effects and assume a parity-even power spectrum
\begin{align}
    P_{\pm}(q^0) = P_T(q^0) ,
\end{align}
where the subscript $T$ denotes the transverse polarisation. For subsequent reference, we define the field–strength power per logarithmic interval, 
associated with the first Lorentz invariant $I_1$, as
\begin{align}
  \mathcal{E}^{2}(\bq) - \mathcal{B}^{2}(\bq)
  \;=\;
  \sum_{\sigma} |\bq|^{3}\,|F_{\bq,\sigma}|^{2},
  \label{eq:curly-E-def}
\end{align}
which carries mass dimension~4.  
In analogy, we define the helicity power per logarithmic interval, 
associated with the second Lorentz invariant $I_2$, as
\begin{equation}
  \mathcal{E}(\bq)\,\mathcal{B}(\bq)
  \;=\;
  \sum_{\sigma} |\bq|^{3}\,
  \big|F_{\bq,\sigma}\,\tilde{F}_{\bq,\sigma}\big|.
  \label{eq:curly-EB-def}
\end{equation}
In what follows, we present two concrete examples of stochastic gauge-field backgrounds and explicitly compute the resulting stochastic pair-production rates.

\subsection{Electromagnetic modes in a cold medium}

In astrophysical plasmas, the presence of a background medium breaks Lorentz invariance, so transverse and longitudinal electromagnetic modes propagate with different dispersion relations. In the present analysis, we restrict our attention to the QED case, 
for which the gauge coupling is given by $g = e$.  In a cold plasma, these take the simple form \cite{Raffelt:1996wa}
\begin{align}
 \omega^2_{\bq,T} = \lvert \bq \rvert^2 + \omega_p^2, \,  \quad \omega^2_{\bq,L} =  \omega_p^2+\mathcal{O}(\mathrm{T}), 
\label{eq:Plasmadisp}
\end{align}
where $\omega(\bq) = q^0$ and $\omega_p$ is the plasma frequency
\begin{equation}
    \omega_p = \sqrt{\frac{n_e e^2}{\varepsilon_0 m_e}},
\end{equation}
defined in terms of the electron number density $n_e$, the elementary charge $e$, the vacuum permittivity $\varepsilon_0$, and the electron mass $m_e$. Note that in \cref{eq:Plasmadisp} we have used $\mathcal{O}(\mathrm{T})$ to denote temperature-dependent corrections. At finite temperature, the Schwinger mechanism becomes thermally assisted: thermal fluctuations populate excited states, effectively lowering the tunnelling barrier and enhancing pair production \cite{Gould:2017fve}. In this work, however, we focus on the cold-plasma regime, where thermal effects are negligible and the pair creation process is governed purely by quantum tunnelling from the vacuum.
In this regime, longitudinal modes correspond to plasma oscillations (Langumir waves),\footnote{Physically, this is a collective density oscillation of electrons relative to ions. Its dispersion relation is given by 
\(\omega(\bq) \approx \omega_p^2 + \frac{3k_B \mathrm{T}}{m_e} |\bq|^2\) for small \(\mathrm{T}\), 
where \(k_B\) is the Boltzmann constant, and \(\mathrm{T}\) is the temperature.} which occur at nearly fixed frequency $\omega_p$, while transverse modes propagate with a dispersion relation resembling that of massive particles. Making use of (\ref{eq:FF--}), we find the scalar invariant
\begin{equation}
   \llangle - \hat{F}_{\mu\nu}(q) \hat{F}^{\mu\nu}(-q)\rrangle = \frac{4VT}{(2\pi)^4} \, \left[\omega_p^2 \delta(\omega-\omega_{\bq,T})\,  P_T(\omega) + \frac{(\lvert \bq\rvert^2-\omega_p^2)^2}{2\omega_p^2} \, \delta(\omega-\omega_p)\, P_L(\omega)  \right].
\end{equation}
Now, if we consider the kinematic constraint $-q^2>4m^2$ and assume $\omega_p>2m$, we find that the contribution of the longitudinal modes is confined to a narrow interval of phase space. On the contrary, the frequency of the transverse modes is only bounded by the cut-off scales, which can lead to arbitrarily large domains of integration. As a result, we restrict our analysis to particle creation arising from transverse electromagnetic modes. Using \cref{eq:curly-E-def} for the transverse case, we have
\begin{equation}
  \mathcal{E}^{2}(\bq) - \mathcal{B}^{2}(\bq)
  \;=\; \Delta_T^2(\bq) \, |\bq|^3 \, P_{T}(\omega_{\bq,T}).
\end{equation}

Equations \eqref{eq:n-omega} and \eqref{eq:spectrum--} allow us to compute the spectral number density produced by the stochastic Schwinger effect for bosons as
\begin{equation}
\llangle n_{\text{EM}}^b({\omega}) \rrangle
= \frac{Te^2Q^2\, \omega^2_p}{3\pi^2} 
\, \left(1 - \frac{4m^2}{\omega^2_p}\right)^{3/2}
\Theta\!\left(\omega_p -  2m\right) \, \omega(\omega^2-\omega_p^2)^{\frac12}\, P_T(\omega).
\label{eq:n-omega-EM}
\end{equation} 
We now compute the vacuum decay rate into a scalar species of mass $m$ in the presence of a cold–plasma background. 
This quantity admits exact analytical solutions  in terms of the Gaussian hypergeometric function ${}_2F_1(a,b;c;z)$. 
The full derivation and closed forms are presented in Appendix \ref{sec:app-compute} (see \cref{eq:app-Hyp---D-EM}). 
For phenomenological clarity, here we report only the simplified expression obtained in the representative regime 
$\Lambda_{\mathrm{IR}}/\omega_p \simeq 1$ and $\Lambda_{\mathrm{UV}}/\omega_p \gg 1$; 
the final result reads
\begin{equation}
\Gamma_\text{EM}^b = \frac{e^2Q^2(\mathcal{E}_*^2 - \mathcal{B}_*^2)}{3\pi^2} \, \mathcal{D}_\text{EM} 
\, \left(1 - \frac{4m^2}{\omega^2_p}\right)^{\frac{3}{2}}
\Theta\!\left(\omega_p -  2m\right),
\label{eq:EMSSch-b}
\end{equation} 
where we used $(\mathcal{E}_*^2 - \mathcal{B}_*^2) \approx\mathcal{C}_T \omega_p^2 \omega_\star^2 $ and $ \mathcal{D}_\text{EM}$ is
\begin{equation}
  \mathcal{D}_\text{EM}   \approx    \begin{cases}
  \frac{1}{3-\delta_T}\left(\frac{\omega_p}{\omega_\star}\right)^{-\delta_T}\left(\frac{\Lambda_\text{UV}}{\omega_p}\right)^{3-\delta_T}  & \delta_T<3\\
  &\\
 \frac{1}{3}\left(\frac{\omega_p}{\omega_\star}\right)^{\gamma_T-2}\,
{}_2F_1\!\left(\frac{3}{2},\,\frac{2-\gamma_T}{2};\,\frac{5}{2};\,1-\frac{\omega_\star^2}{\omega_p^2}\right)  + \ln\left( \frac{\Lambda_\text{UV}}{\omega_p}\right) & \delta_T =3 \\
 &\\
 \frac{1}{3}\left(\frac{\omega_p}{\omega_\star}\right)^{\gamma_T-2}\,
{}_2F_1\!\left(\frac{3}{2},\,\frac{2-\gamma_T}{2};\,\frac{5}{2};\,1-\frac{\omega_\star^2}{\omega_p^2}\right)+\frac{\sqrt{\pi}}{4}\frac{\Gamma\left(\frac{\delta_T-3}{2}\right)}{\Gamma\left(\frac{\delta_T}{2}\right)}\left(\frac{\omega_p}{\omega_\star}\right)^{3-\delta_T} & \delta_T>3
\end{cases}.
\label{eq:DEM}
\end{equation}

We now turn to study the production of charged Dirac fermions with mass $m$. Following the same steps as before and employing the same notation, their spectral number densities and creation rate per unit spacetime volume is related to that of charged bosons by
\begin{equation}
  \llangle n_{\text{EM}}^f({\omega}) \rrangle =  4\left(\frac{\omega_{p}^2+2m^2}{\omega_{p}^2-4m^2}\right) \, \llangle n_{\text{EM}}^b({\omega}) \rrangle , \quad\quad \Gamma_{\text{EM}}^f =  4\left(\frac{\omega_{p}^2+2m^2}{\omega_{p}^2-4m^2}\right) \, \Gamma_{\text{EM}}^b.
\label{eq:EMSSch-f}
\end{equation}
This evidences that fermionic pair creation is larger at equal coupling strength and mass. A concrete example of an astrophysical stochastic electromagnetic background is provided by gamma-ray bursts (GRBs), which emit keV–TeV photons from relativistic outflows following stellar collapse or compact-object mergers. The spectral energy density of GRBs is well described by the empirical Band function \cite{Band1993ApJ}, which yields characteristic spectral indices $\gamma_{T}\approx 0$ and $\delta_{T}\approx 3.5$. adopting GRB-like Band indices and considering $\frac{\omega^2_p}{\omega^2_\star}\lesssim \mathcal{O}(10^{-1})$, we find 
\begin{equation} 
\mathcal{D}_\text{EM} \approx 6.8 \quad \text{for GRBs}.
\end{equation}

Since the pair-production rate is proportional to $\Theta(\omega_p - 2m)$, 
the process is kinematically allowed only when the plasma frequency exceeds twice the particle mass, 
$\omega_p > 2m$. In cold astrophysical environments, however, the plasma frequency typically lies 
well below the keV scale, rendering the production of Standard Model fermions, such as 
electron–positron pairs, impossible. In principle, the stochastic Schwinger mechanism could 
operate efficiently for light (sub-keV) millicharged dark-sector particles, for which the kinematic 
threshold is naturally satisfied. Yet, stringent stellar-cooling bounds impose 
$Q \lesssim 10^{-14}$~\cite{Davidson:2000hf}, which suppresses the corresponding rate to a negligible level 
in cold astrophysical plasmas. The plasmon decay identified in freeze-in dark-matter scenario introduced in \cite{Cora2019} is conceptually related to the stochastic Schwinger effect discussed here: both describe gauge-field energy conversion into pairs once $\omega_p > 2m$, but while plasmons are thermal, quantised in-medium excitations, the stochastic Schwinger effect arises from a real, cold ($\mathrm{T}/m_e\ll 1$), stochastic classical background with a similar kinematic structure.

We conclude this section by commenting on another related but conceptually distinct mechanism, the 
Breit--Wheeler process, $\gamma_1+\gamma_2\!\to\!e^++e^-$~\cite{Breit:1934zz}. 
This process corresponds to perturbative pair creation through photon–photon scattering and 
establishes the kinematic threshold for $\gamma$-ray absorption, 
$E_{\gamma_1}E_{\gamma_2}\!\gtrsim\!(m_e c^2)^2$. 
From a theoretical perspective, the stochastic channel interpolates between the static Schwinger and Breit--Wheeler 
regimes: like the former, the rate is controlled by the field strength, while the stochastic 
temporal structure of the background introduces an effective frequency (or kinematic) threshold 
analogous to that of the Breit--Wheeler process.

\subsection{Dark photon background }
Dark photons are hypothetical massive spin-1 bosons from a hidden \(U(1)'\) sector, 
kinetically mixed with the Standard Model hypercharge 
(see~\cite{Fabbrichesi_2021, Caputo:2021eaa} for reviews). 
In astrophysical and cosmological contexts, stochastic dark-photon fields may form 
with sizeable energy densities, acting as classical sources for charged-particle production and providing a natural setting for the stochastic Schwinger effect \cite{Fedderke_2021, Cline:2024wja, Caputo:2025avc}. A massive vector field of mass $m_{A'}$ propagates three physical degrees of freedom: two transverse and one longitudinal, with dispersion relation
\begin{equation}
    \omega_{\bq, \sigma}^2=|\bq|^2+m_{A'}^2,
\label{Eq:Darkphoton_dispersion}
\end{equation}
which holds both in vacuum and, to excellent approximation in dilute astrophysical media, as plasma-induced modifications are suppressed for weakly mixed dark photons. In such set-ups, the scalar invariant remains positive due to the presence of the mass term,
\begin{equation}
    \llangle- \hat{F}_{\mu\nu}(q) \hat{F}^{\mu\nu}(-q)\rrangle = \frac{4VT}{(2\pi)^4}m_{A'}^2  \, \left[ P_T(\omega) + \frac{m_{A'}^2}{2\omega^2} \, P_L(\omega)  \right] \, \delta(\omega-\sqrt{|\bq|^2+m_{A'}^2}) .
\label{Eq:FF_darkphoton}
\end{equation}
In the presence of time-dependent charge densities, the longitudinal polarisation couples directly to density fluctuations and dominates over the transverse component. We therefore neglect $P_T$ in our analysis. Using \cref{eq:curly-E-def} for the longitudinal polarisation modes, we have
\begin{align}
 \mathcal{E}^{2}(\bq) - \mathcal{B}^{2}(\bq)
  \;=\; \frac{m^4_{A'}}{|\bq|^2+m^2_{A'}} \, |\bq|^3 \, P_{L}(\omega_{\bq,L}).
\end{align}
Using \cref{eq:n-omega}, we find the contribution of the longitudinal mode to the spectral number density to be
\begin{equation}
\llangle n^b_{A'}(\omega)\rrangle=\frac{Tg^2Q^2m_{A'}^4}{6\pi^2}\left(1-\frac{4m^2}{m_{A'}^2}\right)^{\frac32} \Theta(m_{A'}-2m)\, (1-\frac{m_{A'}^2}{\omega^2})^{\frac{1}{2}}\,{P_L(\omega)}.
\label{eq:n-omega-A}
\end{equation}
We now evaluate the vacuum decay rate into a scalar species of mass $m$ in the presence of a cold–plasma background. 
This quantity also admits an exact analytical expression in terms of the Gaussian hypergeometric functions; the detailed derivation and the resulting closed-form expressions can be found in Appendix \ref{sec:app-compute} (see \cref{eq:app-Hyp---D-A}).

As before, we are able to find the total decay rate by into scalars of mass $m$ by integrating over the physical frequency space, under the assumptions 
$\Lambda_{\mathrm{IR}}/m_{A'} \simeq 1$ and $\Lambda_{\mathrm{UV}}/m_{A'} \gg 1$,
\begin{equation}
    \Gamma_{A'}^b=\frac{g^2Q^2(\mathcal{E}_\star^{2} - \mathcal{B}_\star^{2})}{6\pi^2}\,  \mathcal{D}_{A'}\,\left(1-\frac{4m^2}{m_{A'}^2}\right)^{\frac32}\,\Theta(m_{A'}-2m)\,
\label{eq:ASSch-b}
\end{equation}
where $\mathcal{E}_\star^{2} - \mathcal{B}_\star^{2}\approx\mathcal{C}_L m_{A'}^4 $ and $ \mathcal{D}_{A'}$ can be approximated as
\begin{equation}
  \mathcal{D}_{A'}   \approx    \begin{cases}
  \frac{1}{1-\delta_L}\left(\frac{m_{A'}}{\omega_\star}\right)^{-(\delta_L+2)}\left(\frac{\Lambda_\text{UV}}{m_{A'}}\right)^{1-\delta_L}  & \delta_L<1\\
  &\\
 \frac{1}{3}\left(\frac{m_{A'}}{\omega_\star}\right)^{\gamma_L-4}\,
{}_2F_1\!\left(\frac{3}{2},\,\frac{4-\gamma_T}{2};\,\frac{5}{2};\,1-\frac{\omega_\star^2}{m_{A'}^2}\right) + \ln\left( \frac{\Lambda_\text{UV}}{m_{A'}}\right) & \delta_L =1 \\
 &\\
 \frac{1}{3}\left(\frac{m_{A'}}{\omega_\star}\right)^{\gamma_L-4}\,
{}_2F_1\!\left(\frac{3}{2},\,\frac{4-\gamma_T}{2};\,\frac{5}{2};\,1-\frac{\omega_\star^2}{m_{A'}^2}\right)+\frac{\sqrt{\pi}}{4}\frac{\Gamma\left(\frac{\delta_L-1}{2}\right)}{\Gamma\left(\frac{\delta_L+2}{2}\right)}\left(\frac{m_{A'}}{\omega_\star}\right)^{1-\delta_L} & \delta_L>1
    \end{cases}.
\end{equation}
In analogy with \cref{eq:EMSSch-f}, the corresponding fermionic spectral density and production rate follow as
\begin{equation}
  \llangle n_{{A'}}^f({\omega}) \rrangle =  4\left(\frac{m_{A'}^2+2m^2}{m_{A'}^2-4m^2}\right) \, \llangle n_{A'}^b({\omega}) \rrangle , \quad\quad \Gamma_{{A'}}^f =  4\left(\frac{m_{A'}^2+2m^2}{m_{A'}^2-4m^2}\right) \, \Gamma_{{A'}}^b.
\label{eq:ASSch-f}
\end{equation}   
As the rate is proportional to $\Theta(m_{A'} - 2m)$, pair production becomes possible only once the dark photon is heavy enough to overcome the kinematic threshold, $m_{A'} > 2m$. 

The charged particles coupled to the dark photon may belong either to a dark sector, consisting of 
fermionic or bosonic states, or to the Standard Model, in which case they carry a small effective 
(millicharge) coupling $gQ = \varepsilon e$. In the dark-sector case, the coupling $gQ$ can naturally 
be of order unity, allowing a stochastic dark-photon background with mass $m_{A'}$ to produce pairs 
with masses up to $m = m_{A'}/2$. For SM particles, however, current experimental and 
astrophysical constraints on $\varepsilon$ impose stringent limits on such interactions. Collider 
and electroweak precision measurements yield $\varepsilon \lesssim 10^{-2}\,\text{–}\,10^{-3}$ for 
$m_{A'}$ in the GeV range, while at higher masses, $m_{A'} \gtrsim 100~\mathrm{GeV}$, these bounds 
become less restrictive~\cite{Fabbrichesi_2021}.


\section{Stochastic Schwinger in Axion–Gauge Field Reheating}\label{Sec:Phenomenology_II}
In this section we turn to {non-stationary} backgrounds, where time-translation invariance is broken and the fields evolve on observationally relevant timescales.  Here we employ the short-time Fourier framework, \cref{eq:A-non-stationary-mode}, and the Gaussian-windowed unequal-time correlators, \cref{eq:n-omega-non-stationary,eq:n-omega-fermion-non-stationary}, to capture temporally localised production. This treatment offers a clear bridge from first principles to potential signatures of the stochastic Schwinger mechanism in time-dependent environments. As a concrete example of such a stochastic gauge-field background, 
we consider axion--QED interactions during reheating.  At the end of inflation, the Universe enters the reheating phase, during which matter fields evolve atop the coherently oscillating background of the inflaton field. Considering the axion-QED interaction, the gauge fields are transiently amplified by the axion--gauge coupling. This stochastic gauge field background has been computed through lattice simulations 
in \cite{Adshead:2015pva, Adshead:2023mvt}. In what follows, we adopt the standard treatment commonly used in the reheating literature (see  \cite{Lozanov:2019jxc} for a comprehensive review). 

Axion dynamics during preheating can be described by a damped oscillator equation
\begin{align}
\ddot\chi + (3H + \Gamma_{\chi})\dot\chi + m_\chi^2\chi = 0,
\end{align}
where $\Gamma_{\chi}$ denotes the decay rate of the inflaton into other particles, and $m_\chi$ is the axion mass.  Since we neglect the expansion of the Universe in this work, our current analysis is limited to scenarios where the inflaton decay rate is much bigger than the Hubble, $\Gamma_{\chi} \gg H$, in time scales also shorter than the Hubble time, $\Delta t< H^{-1}(t)$. In the regime $m_\chi\gg \Gamma_{\chi} \gg H$, the equation simplifies to
\begin{align}
\ddot\chi + m_\chi^2\chi \approx 0,
\end{align}
with the corresponding solution
\begin{align}
\chi(t) \approx \chi_0\cos(m_{\chi}t),
\label{eq:approx-chi}
\end{align}
where $\chi_0$ remains approximately constant over time scales much shorter than the Hubble time. The axion velocity then follows as
\begin{align}
\dot\chi(t) \approx -m_\chi\chi_0 \sin(m_{\chi}t).
\end{align}
Having understood the free dynamics of our theory, we couple our axion-inflaton field to a $U(1)$ gauge field through a Chern-Simons interaction
\begin{equation}
    \mathcal{L}_{\text{int}} = -\frac{\lambda}{4f} \chi F_{\mu\nu} \tilde{F}^{\mu\nu},
\end{equation}
where $f$ is the axion decay constant and $\lambda$ is a dimensionless coupling constant. This interaction allows for gauge field production during reheating in the regime where the Hubble expansion can be neglected, i.e., on time scales $t \ll H^{-1}$.  To compute the stochastic Schwinger effect, here we only focus on the gauge field background generated during preheating, computed within the domain of validity of our framework.\footnote{A related but distinct consequence of particle production from axion–QED interactions is the chiral memory effect \cite{Maleknejad:2023nyh}, which manifests as a lasting imprint on the spin angular momentum of photons on the celestial sphere.}

Consider the Coulomb gauge ($A_0 = 0$, $\nabla \cdot \vec{A} = 0$), and use the Fourier-decomposed in terms of three-momentum which is the standard notation in cosmology \cite{Weinberg:2008zzc} 
 \begin{align}
   \hat{A}_{\mu}(x) = \sum_{\sigma} \, \int d^3q  \, e^{i\bq.\bx} \Big[ \, A_{\sigma}(t,\bq) e^{\sigma}_{\mu}(\hat{\bq}) \, \hat{\alpha}_{\bq,\sigma} + c.c. \Big],
\end{align}
where the circular polarisation vectors $e^\pm_{\mu}(\hat{\bq})$ are
eigenvectors of rotations about $\hat{\bq}$, i.e.
\begin{equation}
  \hat{\bq} \times {\be}^\pm(\hat{\bq}) \;=\; \mp\, i\, {\be}^\pm(\hat{\bq}),
  \qquad
  \hat{\bq}\cdot {\be}^\pm(\hat{\bq}) = 0,
  \qquad
  {\be}^+(\hat{\bq}) = {\be}^{-\,*}(\hat{\bq}).
\end{equation}
The transverse photon modes satisfy 
\begin{equation}
    \ddot{A}_\pm(t,\bq) + \left(|\bq|^2 \mp |\bq|\frac{\lambda}{f} \dot{\chi}(t) \right) A_\pm(t,\bq) = 0.
\end{equation}
Note that the axion does not generate the longitudinal polarisation. Substituting \cref{eq:approx-chi}, the system reduces to a parametrically driven oscillator, expressible as a Mathieu equation \cite{Adshead:2015pva}
\begin{equation}
    \frac{d^2 A_\pm}{dz^2} + \left[\alpha_{\bq} \mp 2\kappa_{\bq}\sin(2z) \right] A_\pm = 0,
\end{equation}
with
\begin{align}
z = \frac{m_{\chi}t}{2}, \quad \alpha_{\bq} = \frac{4|\bq|^2}{m^2_{\chi}}, \quad \kappa_{\bq} = \frac{2|\bq| \lambda \chi_0}{m_{\chi}f}.    
\end{align}
Here, $\alpha_{\bq}$ encodes the ratio of the gauge to axion oscillation frequency and $\kappa_{\bq}$ controls the strength of the modulation. Solutions to such a system exhibit both tachyonic instability and parametric resonance, leading to exponential growth of gauge degrees of freedom. Over multiple oscillations of the axion, both transverse gauge polarisations are efficiently excited, leading to a nearly unpolarised spectrum. Depending on the size of $\kappa_{\bq}$, we classify the resonant behaviour into the broad  $(\kappa_{\bq}\gg1)$ and narrow $(\kappa_{\bq}\lesssim1)$ bands \cite{Lozanov:2019jxc}. Within the resonance band, solutions obey Floquet behaviour,
\begin{equation}
   A_{\pm}(t,\bq) \approx \frac{c(t,\bq)}{\sqrt{2\lvert \bq \rvert}}\  e^{\mu_{\bq} t}, \qquad \mu_{\bq}t={\kappa_{\bq} z} \sim \xi|\bq|t,
\end{equation}
where $c(t,\bq)$ is a dimensionless periodic function, $\xi = \tfrac{\lambda \chi_{0}}{f}$, and $\mu_{\bq}$ is the Floquet exponent.  The effective instability window is given by
\begin{align}
 1<  \frac{|\bq|}{H} \lesssim \frac{m_{\chi}}{H} \, \xi,
 \label{eq:inst}
\end{align}
where the infrared cut-off corresponds to modes residing within the Hubble horizon. In this work, we focus on early-time dynamics with $\mu_{\bq} t \ll 1$, a regime that captures the most relevant phenomenology while avoiding backreaction. Within the Gaussian-window STFT framework in~\eqref{eq:A-non-stationary-mode}, this corresponds to imposing $\sigma \lesssim \mu_{\bq}^{-1}$.  
As discussed in \cref{sec:non-stationary}, in this context the window width is naturally momentum-dependent, scaling with the oscillation timescale of each mode, $\sigma_{\bq} \sim \omega_{\bq}^{-1}$, in close analogy with adaptive wavelet methods.
From \cref{eq:A-non-stationary,eq:A-non-stationary-mode}, together with \cref{eq:app-gaussian-int} and the identification $\sigma \simeq \mu_{\bq}^{-1}$, the gauge mode functions simplify to  
\begin{equation}
    A_{\pm}(q;t) \approx 
    \frac{1}{2\sqrt{\pi}} \,
    \frac{c_{\pm}}{\xi |\bq|^{3/2}}\,
    \exp\!\left[\tfrac{1}{2}\Big(1 - \tfrac{(q^0)^2}{\xi^2|\bq|^2}\Big) 
    + i\,\tfrac{q^0}{\xi |\bq|} \right] \, e^{\xi |\bq| t} \, e^{iq^0 t}.
\end{equation}
 As we are working in the small time domain, we can approximate the periodic function as being nearly constant, $ c(t,\bq)\approx  1$. 
Using the above, \cref{eq:app-ff}, \cref{eq:app-fff}, and in the early-time regime, i.e. $\xi \lvert \bq \rvert t \ll 1$, we find
\begin{align}
  \llangle - \hat{F}_{\mu\nu}(q^{\mu};t) \hat{F}^{\mu\nu}(-q^{\mu};t) \rrangle  
    &\approx \, \frac{V}{2\pi}
   \,
    \frac{(-q^2)  }{\xi^2 |\bq|^{3}}\,
    \exp\!\left[1 - \tfrac{(q^0)^2}{\xi^2|\bq|^2}\right]+\mathcal{O}{(\xi \lvert \bq \rvert t )}.
\end{align}
Using \cref{eq:curly-EB-def}, the helicity power per logarithmic interval for the axion–QED system can be computed as
\begin{align}
\mathcal{E}(\bq)\!\cdot\!\mathcal{B}(\bq)= (\bq\cdot\bq)^2\, |\xi|.
\end{align}

Starting from the above, we use \cref{eq:Gamma-Boson} to compute the rate of created boson pairs with mass $m$. That gives the number density of the generated pairs
\begin{align}
\begin{split}
    n^b_{\chi} = \int_{H}^{q_{\text{max}}} \, d|\bq| \, \,  n^b_{\chi}(|\bq |), 
    \label{eq:a-QED}
\end{split}
\end{align}
where $ n^b_{\chi}(|\bq |)$ is the 
spectral number density given as 
\begin{equation}
   \llangle n^b_{\chi}(|\bq |) \rrangle \approx \frac{2.7 g^2 \, Q^2}{6\pi \, \xi^2} \, |\bq|^2 \, \int_{\sqrt{1+{4m^2}/{|\bq|^2}}}^{\infty} dy  \left(1-\frac{4m^2/|\bq|^2}{y^2-1}\right)^{\frac{3}{2}}{(y^2-1)} \,  \, e^{ - y^2/\xi^2},
    \label{eq:a-QED--}
\end{equation}
where $y = \omega / |\mathbf{q}|$, with $e \simeq 2.7$ and the maximal momentum 
$q_{\text{max}} = m_\chi \xi$ set by the instability window in~\cref{eq:inst}.
 The above expression does not admit a closed-form analytic solution. 
However, we were able to approximate its behaviour using complementary error functions
as detailed in \cref{sec-app-a-QED} of this work. The explicit form of the spectral number density is presented in \cref{eq:spectraltotal}. 
Its contribution remains negligible for small values of \(|\bq|\) and becomes relevant for \(|{\bq}| \gtrsim 2m/\xi\), 
where it takes the following form 
\begin{equation}
   \llangle n^b_{\chi}(|\bq|) \rrangle \approx \frac{9  g^2 \,Q^2 }{80\sqrt{\pi}} \, \xi \, |\bq|^2 \,  \mathcal{D}_{\chi}\, \Theta(m_\chi\xi^2-2m), 
\label{eq:ab-QED---p}
\end{equation}
where the parameter $\mathcal{D}_\chi\in(0.4, 1)$ is defined as
\begin{equation}
    \mathcal{D}_{\chi}(\xi) \equiv 
    \left[
        \frac{2}{\sqrt{\pi}\,\xi} e^{-1/\xi^{2}}
        + \frac{1}{\xi^{2}}\left(\xi^{2} - 2\right)
        \mathrm{erfc}\!\left(\frac{1}{\xi}\right)\, \Theta(m_\chi\xi^2-2m)
    \right].
\end{equation}
By performing the momentum integral, we obtain the number density given in \cref{eq:ab-QED---o-}. 
In the regime of interest, \(\xi \gtrsim 1\), it becomes 
\begin{equation}
    n^b_{\chi} \approx \frac{3 g^2 \,Q^2 }{80 \sqrt{\pi}} \, m_{\chi}^3 \, \xi^4 \, \mathcal{D}_{\chi}\, \Theta(m_\chi\xi^2-2m).
\end{equation}
Finally, we find the background decay rate to bosons as (see \cref{eq:app-Gamma-chi-QED-b})
\begin{equation}
   \Gamma^b_{\chi} \approx \frac{0.9 }{32\sqrt{\pi}}  \,  g^2 \,Q^2\, (\mathcal{E}\!\cdot\!\mathcal{B})_{\text{max}} \, \mathcal{D}_{\chi}  \, \Theta(m_\chi\xi^2-2m) ,
\label{eq:ab-QED---o-decay}
\end{equation}  
in which     $(\mathcal{E}\!\cdot\!\mathcal{B})_{\text{max}} = m_\chi^4\xi^5$ denotes the value of $\mathcal{E}(\bq)\!\cdot\!\mathcal{B}(\bq)$ evaluated at $|\bq| = m_{\chi}\xi $. Similarly, for the fermionic case, the pair production can be computed. The details of the calculation are presented in \cref{sec-app-a-QED}, and we report here only the final result.
The  corresponding fermionic spectral density and production rate are found to be four times larger than that of bosons with the same mass and charge, i.e.
\begin{equation}
\llangle n^f_{\chi}(|\bq|) \rrangle \approx 4 n^b_{\chi}(|\bq|),  \quad \quad \Gamma^f_\chi \approx 4 \, \Gamma^b_\chi.
\label{eq:ab-QED---hf}
\end{equation}
The energy density of the produced particles can be written as $\rho = m\,n_{\chi}$. 
Our analysis was carried out in the regime where backreaction on the background field is neglected. 
To verify the consistency of this assumption, we compared the produced energy density with that of the background gauge field, 
$\rho_A = \tfrac{1}{2}(E^2 + B^2) \simeq \tfrac{\xi^4}{16}(1 + \xi^2)m_\chi^4$. 
The resulting condition, $\frac{m_\chi}{m} \gg \frac{3}{5\sqrt{\pi}}\,\frac{g^2 Q^2 \mathcal{D}_\chi}{1 + \xi^2}$,
is comfortably satisfied in the parameter regime $m_\chi \gg m$, confirming that backreaction effects are indeed subleading and the approximation self-consistent.

\subsection*{Axion--gauge fields in de Sitter and the Schwinger effect:}
 To place the above discussion in a broader context, let us conclude this section by connecting it to 
related developments in de~Sitter space, which provide a natural extension of the present analysis. 
The static Schwinger mechanism in 4d de~Sitter space was first examined in the seminal 
work~\cite{Kobayashi:2014zza}, which presented a refined QFT analysis of scalar QED and showed that 
vacuum pair production could, in principle, jeopardise the stability of the quasi–de~Sitter background. 
A subsequent comprehensive study~\cite{Bavarsad:2017oyv} further advanced this framework by incorporating 
constant parallel electric and magnetic fields, offering deeper insight into the interplay between 
electromagnetic and gravitational pair production. These works provided valuable physical understanding 
and established an important reference point for later studies. Nonetheless, both relied on the simplifying 
assumption of a constant, homogeneous electric field, an approximation that, while conceptually illuminating, 
neglects backreaction and is inconsistent with the gauge-field dynamics in de~Sitter spacetime, where the 
field redshifts as radiation. Moreover, such a static $U(1)$ background explicitly breaks the spatial 
isotropy of the cosmological geometry. 

Both issues were subsequently addressed in~\cite{Lozanov:2018kpk, Mirzagholi:2019jeb, Maleknejad:2019hdr}, which demonstrated that axion--inflation naturally supports a quasi-static electromagnetic field background, while the isotropic gauge field configuration introduced in~\cite{Maleknejad:2011sq,Maleknejad:2011jw} provides a self-consistent realisation that preserves cosmological isotropy. These developments have opened a new line of research into the static Schwinger effect in axion--inflation models, where the decay rate derived for constant electromagnetic backgrounds in~\cite{Bavarsad:2017oyv} has been applied to the axion–gauge field system (see, e.g.,~\cite{Domcke:2021fee, vonEckardstein:2024tix,Iarygina:2025ncl}). In contrast, recent lattice simulations for the Abelian case~\cite{Caravano:2021bfn,Figueroa:2023oxc,Sharma:2024nfu} reveal that the gauge field generated during inflation evolves into a highly inhomogeneous and intrinsically stochastic configuration. This observation suggests that the stochastic Schwinger mechanism introduced in this work may not only emerge naturally during gauge reheating, but could also represent a generic, and perhaps unavoidable, feature of axion inflation.

 
\section{Static vs.\ Stochastic Schwinger Effects}\label{sec:static-vs-stochastic}

In this section, we assess the efficiency of the stochastic Schwinger mechanism relative to the conventional static-field result. 
To this end, we first use \cref{eq:EMSSch-b,eq:ASSch-b} as representative examples of stochastic decay channels relevant to high-energy astrophysical environments, and compare their rates with the static Schwinger case in \cref{eq:Schwinger-scalar}. 
The corresponding ratios are
\begin{align}
    \frac{\Gamma_{\text{EM}}}{\Gamma_{\text{static}}}
    &\sim 
    \mathcal{D}_{\text{EM}}
    \left(1-\frac{4m^2}{\omega_p^2}\right)^{\!\frac{3}{2}}
    \exp\!\left(\frac{\pi m^2}{g Q E}\right)
    \Theta(\omega_p - 2m),
    \label{eq:compareEM}\\[6pt]
    \frac{\Gamma_{A'}}{\Gamma_{\text{static}}}
    &\sim 
    \mathcal{D}_{A'}
    \left(1-\frac{4m^2}{m_{A'}^2}\right)^{\!\frac{3}{2}}
    \exp\!\left(\frac{\pi m^2}{g Q E}\right)
    \Theta(m_{A'} - 2m),
    \label{eq:compareA}
\end{align}
where we have taken $\mathcal{E}_*^2 - \mathcal{B}_*^2 \simeq E^2$ and truncated the static Schwinger series to its leading ($n=1$) term. 
As noted earlier, the $\Theta$-functions enforce the strict kinematic thresholds for pair creation. 
Equations~\eqref{eq:compareEM} and~\eqref{eq:compareA} demonstrate that the stochastic channel is parametrically enhanced by at least a factor of $\exp(\pi m^2 / g Q E)$. 
Consequently, whenever the background field lies below the critical Schwinger value, $E_c \equiv m^2/(g Q)$, stochastic pair production overwhelmingly dominates. 
Conceptually, the stochastic Schwinger effect occupies an intermediate regime between the static Schwinger and Breit--Wheeler processes: 
like the former, the rate depends on the field strength, yet—owing to the field’s random temporal structure—it also exhibits a frequency threshold reminiscent of the latter.

\medskip

Next, we turn to the non-stationary axion–gauge-field reheating case, \cref{eq:ab-QED---o-decay}, 
in the regime where cosmic expansion can be neglected, as discussed in detail in \cref{Sec:Phenomenology_II}. 
The corresponding ratio can be evaluated and compared with the $\mathbf{E}\!\parallel\!\mathbf{B}$ static Schwinger effect of \cref{eq:EH-scalar}, yielding
\begin{align}
    \frac{\Gamma_{\chi}}{\Gamma_{\text{static}}}
    &\sim 
    \mathcal{D}_{\chi}\,
    \sinh\!\left(\frac{\pi}{\xi}\right)
    \Theta(m_\chi \xi^2 - 2m),
    \label{eq:comparechi}
\end{align}
where we have assumed $E = \mathcal{E}_{\text{max}}$, $B = \mathcal{B}_{\text{max}}$, and $m^2 \ll g Q \xi^3 m_\chi^2$. 
In this case, corresponding to axion–gauge reheating, the stochastic-to-static ratio is generically of order unity. 

In situations where the electromagnetic stochastic component is superimposed on a homogeneous background, we can decompose the gauge potential as
\begin{equation}
\hat{A}_\mu(x) = \bar{A}_\mu(x) + \delta \hat{A}_\mu(x), \qquad 
\langle \delta \hat{A}_\mu(x) \rangle_s = 0,
\end{equation}
where $\bar{A}_\mu(x)$ generates a static electromagnetic field. The background then contributes the usual non-perturbative Schwinger term, while the fluctuations provide an additional stochastic correction set by their two-point correlator. Since the two sectors arise from independent parts of the gauge field, their contributions to the decay exponent add linearly,
\begin{equation}
    \Gamma_{\text{total}}=\Gamma_\text{static}+\Gamma_\text{stochastic}.
\end{equation}
In this regime, the two sectors remain separable at leading order in $\llangle F_{\mu\nu}F^{\mu\nu}\rrangle $, so the static background does not modify the stochastic contribution. This contrasts with dynamically assisted case with $\llangle \delta F_{\mu\nu}(x)\rrangle \neq 0$ with
\begin{align}
F_{\mu\nu}F^{\mu\nu} = (\bar{F}_{\mu\nu}+\delta F_{\mu\nu}(x))(\bar{F}^{\mu\nu}+\delta F^{\mu\nu}(x))  \supset 2\bar{F}^{\mu\nu} \delta F_{\mu\nu}(x),
\end{align}
 where such interference between the background and fluctuations can occur.
 
To put this discussion into perspective, we sharpen the formulation and scope of the stochastic description by distinguishing between deterministic fast oscillations and ensemble stochasticity, which play distinct roles in the Schwinger process.

\begin{itemize}[leftmargin=1em]  
    \item \textbf{Emergent stochasticity from deterministic fast oscillations:} consider a deterministic background field 
    $A_{\mu}(x,t)=\delta A_{\mu}(x,t)$ with rapid spacetime modulations 
    characterized by oscillation scales $(\ell_{\rm osc},\tau_{\rm osc})$.
    When these satisfy
    \begin{align}
   \ell_{\rm osc} \ll \ell_{\rm Sch},
        \qquad
        \tau_{\rm osc} \ll \tau_{\rm Sch},      
    \end{align}
    where $(\ell_{\rm Sch} = \frac{mc^2}{eE},\tau_{\rm Sch}=\ell_{\rm Sch}/c)$ are the Schwinger tunneling scales, the field averages over many oscillations during a tunneling event. A coarse-graining over intermediate scales then induces an effective stochastic field with covariance
    \begin{align}
        G^{\rm eff}_{\mu\nu}(x)
        = \overline{\delta A_{\mu}(x_{0})\,\delta A_{\nu}(x_{0}+x)},
    \end{align}
    even though the underlying field is fully deterministic. In this regime the Schwinger process responds only to these coarse-grained statistics, and the stochastic formalism applies without requiring any intrinsic randomness of the original field.
\item \textbf{Ensemble (macroscopic) stochasticity:} Cosmological electromagnetic fields are stochastic in the measure-theoretic sense: the
background field is drawn from an ensemble $\{ A(x) \}$ with well-defined correlation
functions,
\begin{align}
 \langle A_{\mu}(x)\, A_{\nu}(x') \rangle_s = G_{\mu\nu}(x-x'),   
\end{align}
even though each realization is slowly varying in space and time. In this situation, the
pair-production rate is obtained from the \emph{ensemble average} of the Schwinger
functional,
\begin{align}
\langle \Gamma \rangle_s 
= \int \mathcal{D}A \, P[A] \, \Gamma[A],
\end{align}
where $P[A]$ is the probability measure characterizing the stochastic field (regardless of $\tau_{\rm osc}$ relative to $\tau_{\rm Sch}$);
instead, one performs an average over realizations rather than over rapid microscopic fluctuations. In other words, the Schwinger process is quasi-local in a single realization, but the physically relevant observable in cosmology is the expectation value $\langle \Gamma \rangle_s$. A closely related situation occurs in certain astrophysical environments where the electromagnetic field exhibits strong spatial or temporal variability across many coherence domains. If the physical system (or the observer) effectively samples a large number of such statistically independent patches, then the appropriate prediction is again obtained by averaging over the ensemble of coarse-grained field configurations.
\end{itemize}
In practice, for astrophysical and cosmological electromagnetic fields, it is predominantly the second situation that applies: the fields vary slowly on microscopic scales but are statistically described by an ensemble distribution across many coherence domains, so the physically meaningful prediction is the ensemble-averaged Schwinger rate.

\section{Conclusions \label{Sec:conclusion}}

The Schwinger effect, describing quantum pair creation in strong electromagnetic fields, has been extensively studied in the context of deterministic background fields—most notably in constant electric fields or in spatially inhomogeneous yet fully deterministic configurations such as high-intensity laser pulses. 
In this work, we extend the mechanism beyond these deterministic settings to encompass {stochastic} background fields. 
A related stochastic particle-production mechanism was recently identified in the gravitational context~\cite{Maleknejad:2024ybn,Maleknejad:2024hoz}, where cosmic perturbations induce particle creation through purely quantum effects. 
Here, we develop the corresponding framework for generic Abelian gauge theories interacting with charged particles, encompassing but not limited to quantum electrodynamics, and derive the effective action governing pair creation in stochastic backgrounds. 
This general formulation yields closed analytical expressions for the pair-production probability of scalar and fermionic particles and can be readily applied to scenarios beyond the Standard Model that involve Abelian gauge sectors. Our analysis was performed in flat spacetime, corresponding to the regime in which the expansion of the Universe can be safely neglected during the particle-production process. 


We have analysed pair production in both {stationary} and {non-stationary} stochastic gauge backgrounds. 
Stationary fields, with a well-defined frequency decomposition, yield scalar and fermionic spectra given by \cref{eq:n-omega,eq:n-omega-fermion-}; they exhibit the kinematic threshold $\omega^2>|\bq|^2+4m^2$, scale quadratically with the coupling ($\propto g^2Q^2$), and are linear in the stochastic power $\llangle -F_{\mu\nu}F^{\mu\nu}\rrangle$, which encodes the spectral strength of gauge fluctuations. 
We then extended the framework to transient, non-stationary settings—where time-translation invariance is broken. 
There, we employ the short-time Fourier transform (STFT), cf.~\cref{eq:A-non-stationary-mode}, with a Gaussian window ($\sigma\!\sim\!\Delta t$, and mode-adaptive $\sigma_{\bq}\!\sim\!\omega_{\bq}^{-1}$ when appropriate), and show that the spectra are governed by the Gaussian-windowed {unequal-time} field-strength correlator, \cref{eq:n-omega-non-stationary,eq:n-omega-fermion-non-stationary}, smoothly reducing to the stationary limit when the correlator depends only on $t'-t''$. 

In Sections \ref{Sec:Phenomenology_I} and \ref{Sec:Phenomenology_II}, we illustrate this mechanism through representative phenomenological examples in the stationary and non-stationary regimes, respectively.
For the stationary case, we consider two scenarios relevant to high-energy astrophysics  \cref{Sec:Phenomenology_I}. The first involves electromagnetic modes in a cold medium with plasma frequency $\omega_p$, where the background is treated as a stochastic electromagnetic field. The corresponding background decay into charged pairs is given by \cref{eq:EMSSch-b} for bosons and \cref{eq:EMSSch-f} for fermions, with a rate proportional to $e^2$ and to the electromagnetic invariant $\mathcal{E}_*^2 - \mathcal{B}_*^2$, evaluated at the peak of the stochastic spectrum. The process is kinematically allowed only when the plasma frequency exceeds twice the particle mass, $\omega_p > 2m$. However, since the plasma frequency in typical astrophysical environments lies well below the keV scale, the resulting stochastic decay in the Standard Model sector is negligible compared to the quantum Breit–Wheeler process. Next, we consider stochastic dark-photon fields acting as sources for the stochastic Schwinger effect. 
The corresponding background decay rates into bosonic and fermionic pairs are presented in 
\cref{eq:ASSch-b} and~\cref{eq:ASSch-f}, respectively. The decay exhibits similar qualitative features 
to the cold-plasma case; however, it is kinematically possible only when the dark-photon mass exceeds 
twice the particle mass, $m_{A'} > 2m$. This condition makes the process significantly more efficient 
than in the electromagnetic case for the production of light dark-sector particles.

In \cref{Sec:Phenomenology_II}, we examined non-stationary, transient backgrounds and, as an illustrative case, considered axion–QED interactions during reheating, where the axion–gauge coupling transiently amplifies stochastic gauge fields. Focusing on the early-time regime—where backreaction remains negligible—and neglecting cosmic expansion, our analysis applies to scenarios in which the inflaton decay rate greatly exceeds the Hubble rate and to timescales shorter than a Hubble time. The bosonic spectral density and production rate are given in \cref{eq:ab-QED---p} and \cref{eq:ab-QED---o-decay}, while the corresponding fermionic expressions are presented in \cref{eq:ab-QED---hf}. The resulting stochastic Schwinger effect scales as $\mathcal{E}(\bq)\!\cdot\!\mathcal{B}(\bq)\big|_{m_\chi\xi} = m_{\chi}^4\xi^5$ and becomes significant when the mass of pairs is less than $ \tfrac{1}{2}\,\xi^2 m_\chi$, where $\xi = \lambda \chi_0 / f$ and $m_\chi$ is the axion mass. This indicates that the stochastic Schwinger mechanism introduced in this work naturally arises during gauge reheating and may play a dynamically important role.

Extending the present analysis to expanding geometries is a natural next step, allowing one to go beyond the quasi-flat regime ($\Gamma\gg H$) explored here and to develop a unified framework applicable to cosmological backgrounds. Furthermore, while our analytical results capture the key qualitative and parametric features of the phenomenon, a fully quantitative understanding, particularly in regimes where backreaction and non-linear effects become significant, will require numerical simulations beyond our current approximations. We leave such investigations to future work.

\section*{Acknowledgments}
We would like to thank Luis Álvarez-Gaumé and Andrew J. Tolley for the insightful discussions that inspired this work. LVGC is grateful for the hospitality of Center for Quantum Fields and Gravity at the University of Swansea.  LVGC and AM are supported by the Royal Society University Research Fellowship, Grant No. RE22432.
\appendix

\section{Mathematical tools}

We collect here useful analytical identities and asymptotic/integral representations used throughout the paper. \newline
{\textbf{Proper-Time Representation of Logarithms and Determinants:}} The natural logarithm admits the proper-time integral representation
\begin{align}
\ln A = \int_{0}^{\infty} \frac{ds}{s}\,\Big[e^{-s} - e^{-sA}\Big],
\qquad A>0 .
\end{align}
This identity allows one to express the determinant of an operator $O$ in terms of its heat kernel. Formally, for a complex scalar field we have
\begin{align}
\ln \det O_b
= \tr \ln O_b
= - \tr \int_{0}^{\infty} \frac{ds}{s}\, e^{-s O_b} .
\label{app:eq-In-Det}
\end{align}
Notice for a real scalar field there is an extra factor of $\frac12$. For fermionic operators, the functional integral over Grassmann variables gives
\begin{align}
\ln \det O_f
= + \tr \int_{0}^{\infty} \frac{ds}{s}\, e^{-s O_f},
\end{align}
in contrast to the bosonic case \eqref{app:eq-In-Det}. This sign flip is the direct consequence of Grassmann integration rules.
\newline
\textbf{Gaussian function:} In the two asymptotic regimes, the Gaussian function admits the following limits. 
As $\sigma \to 0^{+}$, it converges to the Dirac delta distribution,
\begin{align}
\delta(x-a) \;=\; \lim_{\sigma \to 0^{+}} 
\frac{1}{\sqrt{2\pi}\,\sigma}\,
\exp\!\left[-\frac{(x-a)^{2}}{2\sigma^{2}}\right],
\end{align}
while in the opposite limit $\sigma \to \infty$, the Gaussian tends to unity,
\begin{align}
\lim_{\sigma \to \infty} 
\exp\!\left[-\frac{(x-a)^{2}}{2\sigma^{2}}\right] = 1.
\label{eq:app-Gaussian-1}
\end{align}
The basic Gaussian integrals (for $a>0$) are
\begin{align}
\int_{-\infty}^{\infty} e^{-a x^{2}}\,dx = \sqrt{\frac{\pi}{a}},
\qquad
\int_{-\infty}^{\infty} e^{-a x^{2} + b x + c}\,dx
= \sqrt{\frac{\pi}{a}}\, e^{\frac{b^{2}}{4a}+c}.
\label{eq:app-gaussian-int}
\end{align}
\newline
\textbf{Gaussian error function:}  The error function and its complementary form are defined respectively as
\begin{align}
\mathrm{erf}(x) &= \frac{2}{\sqrt{\pi}}\int_0^x e^{-t^2}\,dt, 
\\[4pt]
\mathrm{erfc}(x) &= 1 - \mathrm{erf}(x)
= \frac{2}{\sqrt{\pi}}\int_x^{\infty} e^{-t^2}\,dt.
\label{eq:app-error-c}
\end{align}
Its integral is given by
\begin{equation}
    \int dx\, \text{erf}(x) = x\, \text{erf}(x)+ \frac{1}{\sqrt{\pi}} e^{-x^2}+C
\end{equation}
\newline
\textbf{Hypergeometric function:} It has the Euler integral representation 
\begin{equation}
{}_2 F_1(a,b;c;x)
=\frac{\Gamma(c)}{\Gamma(b),\Gamma(c-b)}
\int_{0}^{1} t^{b-1}(1-t)^{c-b-1}(1-xt)^{-a}dt.
\qquad \Re(c)>\Re(b)>0.
\end{equation}

The following integral admits a closed form in terms of the Gauss hypergeometric function
\begin{align}
\frac{1}{2}\int t^{1/2}\,(t+b^2)^{\frac{n-1}{2}}\,dt
&=\frac{b^{\,n-1}}{3}\,t^{3/2}\,
{}_2F_1\!\left(\frac{3}{2}, \frac{1-n}{2};\,\frac{5}{2};\,-\frac{t}{b^{2}}\right)
+ C.
\label{eq:app-Hyp---}
\end{align}
One special value we will use is
\begin{equation}
{}_2F_{1}\!\left(\frac32, \frac{1-n}{2};\,\frac{5}{2};\,0\right)  =1, \label{eq:2F1-0} 
\end{equation}
These identities hold for generic parameters by analytic continuation, except at gamma-function poles. For $n=-2$, the integral in \cref{eq:app-Hyp---} evaluates to
\begin{align}
   \int \frac{\sqrt{x^{2}-b^{2}}}{x^{2}}\,dx
= \frac{1}{2}\ln\!\left(\frac{\sqrt{1-b^{2}/x^2}+1}{\sqrt{1-b^{2}/x^2}-1}\right)
-\frac{\sqrt{x^{2}-b^{2}}}{x}+ C\,\qquad (|x|\ge b,\; b>0).
\end{align}


\section{Technical Details of Stochastic Effective Action}\label{App:Calculations}

In this appendix, we provide the explicit computations of the effective action discussed in Sec.~\cref{Sec:Stochastic_Scwhinger_Effect}. We begin by proving \cref{sec:QED-expansion}. The $O(A)$ operator in \cref{eq:mathcal-O} can be written as
\begin{equation}
    O^{\dag}(A) - O(A)= O^{\dag}(A) \Big[ \frac{1}{\hat{p}^2+m^2+i\epsilon} - \frac{1}{\hat{p}^2+m^2-i\epsilon}\Big] O(A).
\end{equation}
In the course of the computation, both the Feynman and retarded propagators appear. For clarity, we briefly recall their definitions and roles. The Feynman propagator arises naturally in time-ordered correlation functions and is defined by
\begin{equation}
   G_F(x-y) = \bra{0} T \phi(x) \phi(y)\ket{0}= \int d^4q\, \frac{-i}{q^2+m^2 - i \epsilon} \, e^{-i q(x-y)},
\end{equation}
where the $i\epsilon$ shifts one pole above and the other one below the real axis. The retarded propagator, in contrast, captures causal response and is given by
\begin{equation}
    G_R(x-y)= \Theta(x^0-y^0) \bra{0}  [\phi(x), \phi(y)]\ket{0}= \int d^4q\, \frac{-i}{q^2+m^2 + i \epsilon_{\mu} q^\mu} \, e^{-i q(x-y)} ,
\end{equation}
in which $\epsilon_{\mu}=(\epsilon, 0,0,0)$ is an infinitesimal positive time-like four-vector that shifts both poles below the real axis. The difference between the Feynman and retarded propagators,
\begin{align}
\begin{split}
    G_{F}(x-y)- G_R(x-y) &= \Delta_-(x-y)\\
    &= -i \int d^4q\, \delta(q^2+m^2)\, \Theta(-q^0) e^{-i q(x-y)}.
\end{split}
\end{align}
isolates the negative-frequency on-shell component of the field, i.e. the retarded component of the Wightman function. The inner product of momentum eigenstates reads
\begin{equation}
   \langle q | q' \rangle = \delta^{(4)}(q'-q).
\end{equation}
The translation operator in momentum space acts as a shift,
\begin{equation}
   \langle q | e^{i k \cdot \hat{x}} | q' \rangle = \langle q | q' + k \rangle .
\end{equation}
In deriving this relation we inserted the resolution of identity,
\begin{equation}
   \hat{\mathbb{I}} = \int d^4q \, |q\rangle \langle q| ,
   \label{Eq:ResolutionUnity_SSE}
\end{equation}
with the overlap between momentum and position eigenstates given by
\begin{equation}
   \langle x | q \rangle = e^{-i q \cdot x}.
\end{equation}

Using the positive- and negative-energy mass-shell delta functions defined in \cref{eq:pm-delta}, their matrix elements in the momentum basis are given by
\begin{equation}
\langle q | \hat{\delta}^{\pm}(\hat{p}) | q' \rangle
= \delta^{(4)}(q-q')  \Theta(\pm q^0) \delta(q^2 + m^2).
\label{eq:rho-pm}
\end{equation}
Using the above, we find
\begin{equation}
    \hat{O}^{\dag}(A) - \hat{O}(A)= -i \hat{O}^{\dag}(A) \Big( \hat{\delta}^+ +\hat{\delta}^- \Big) \hat{O}(A).
    \label{eq:FF}
\end{equation}
The det of the  S-matrix for the retarded propagator is one, i.e.
\begin{align}
\begin{split}
1 = & \exp\left[\text{Tr}\ln\left( \hat{\mathbb{I}}  - \frac{g Q \, \hat{f}(A)}{\hat{p}^2+m^2 - i \epsilon \hat{p}^0}\right)\right]\\
 = & \exp\left[\text{Tr}\ln\left( \hat{\mathbb{I}}  - \frac{ g Q \, \hat{f}(A)}{\hat{p}^2+m^2 - i \epsilon }  + i g Q \hat{f}(A) \, \hat{\delta}^ -  \right)\right] \\
 = & \exp\left[\text{Tr}\ln\left( \left( \hat{\mathbb{I}}  - \frac{g Q\, \hat{f}(A)}{\hat{p}^2+m^2 - i \epsilon } \right) \, (1 + i \hat{O}(A) \, \hat{\delta}^ - ) \right)\right]\\
 =&\exp\left[i\Gamma^b_\text{1-loop}\right]\exp\left[\Tr\ln\left(1+i\hat{O}(A)\hat{\delta}^-\right)\right],
\end{split}
\end{align}
which gives the effective action as
\begin{equation}
  i\Gamma_{\text{1-loop}}^b[A_\mu] = 
  -\text{Tr}\ln\left(  \hat{\mathbb{I}}  + i \hat{O}(A) \, \hat{\delta}^ - \right).
    \label{eq:S-int}
\end{equation}
Now, we obtain
\begin{align}
\begin{split}
    \text{Im}\Gamma^{b}_\text{1-loop}[A_\mu]&=-\Tr\ln\left[(\hat{\mathbb{I}}+i\hat{O}(A)\hat{\delta}^-)(\hat{\mathbb{I}}-i\hat{O}^\dagger(A)\hat{\delta}^-)\right]\\
    & = -\Tr\ln\left[\hat{\mathbb{I}}-i(\hat{O}(A)^\dagger - \hat{O}(A))\hat{\delta^{-}}+\hat{O}(A)\hat{\delta}^-\hat{O}(A)\hat{\delta}^-\right]\\
    &=-\text{Tr}\ln\left[  \hat{\mathbb{I}}  - \hat{O}(A) \, \hat{\delta}^+ \hat{O}(A) \, \hat{\delta}^- \right],
\label{eq-app:QED-expansion}
\end{split}
\end{align}
where to arrive at the last line we have used the identity \cref{eq:FF}.

We now turn to the proof of \cref{sec:QED-expansion}. Starting from \cref{eq-app:QED-expansion} we can expand the above expression perturbatively to order $g^2$ to obtain
\begin{align}
    \begin{split}
        \text{Im}\Gamma_{\text{1-loop}}^b[A_\mu] &=-g^2Q^2\Tr[(2A_\mu\partial^\mu+ \partial^\mu A_\mu)\hat{\delta}^+(2A_\nu\partial^\nu+\partial^\nu A_\nu)\hat{\delta}^-]\\
        &=-g^2 Q^2\int\!\!\!\int\!\!\!\int_{x,p_1,p_2} \bra{p_1}\llangle (2A_\mu(x) \partial^\mu+\p^\mu A_\mu(x))\hat{\delta}^+(x)\\
        & \quad\times \, (2A_\nu(x)\partial^\nu+\p^\nu A_\nu(x))\hat{\delta}^-(x)\rrangle \ket{p_2}\langle{x}\ket{p_1} \, \langle {p_2}\ket{x},    
    \end{split}
\end{align}
where for shorthand, we denote $\int_q \equiv \int d^4 q$. The proof of \cref{eq:Gamma-Boson} now proceeds as follows. We begin by writing
\begin{align}
\bra{p} A_{\mu}(X) \ket{p-q} = \frac{1}{(2\pi)^2} A_{\mu}(q) = \frac{1}{(2\pi)^4} \int d^4 x \, A_{\mu}(x) \, e^{-iq.x},
\end{align}
and making use of \cref{eq:rho-pm}, we find
\begin{align}
\begin{split}
        \text{Im}\Gamma_{\text{1-loop}}&=g^2 Q^2\int\!\!\!\int_{p_1,p_2}  \llangle A_{\mu}(p_1-p_2) A_{\nu}(p_2-p_1)\rrangle \, \delta(p_1^2+m^2) \delta(p_2^2+m^2) \Theta(p_1^0) \Theta(-p_2^0) \nonumber\\
        & \times (2p_{1}^{\nu} - (p_1-p_2)^\nu) (2p_2^{\mu} + (p_1-p_2)^\mu)\, \nonumber\\
&=g^2 Q^2\int d^4q \,  \llangle A_{\mu}(q) A_{\nu}(-q)\rrangle \, \int\!\!\!\int_{p_1,p_2}  (2p_{1}-q)^{\nu} (2p_2+q)^{\mu}\,\delta(p_1^2+m^2)\\
        & \times \delta^{(4)}(q-p_1+p_2)\, \delta(p_2^2+m^2)  \, \Theta(p_1^0) \Theta(-p_2^0).
        \label{eq:W1-int}
\end{split}
\end{align}
It will be convenient to make this explicit 
\begin{equation}
\llangle \hat{A}_{\mu}( q) \hat{A}_{\nu}(-q)\rrangle = \frac{1}{(2\pi)^4} \int d^4q' \,  \int d^4x \,  \llangle \hat{A}_{\mu}( q) \hat{A}_{\nu}(q')\rrangle \, e^{i(q-q').x},
\label{eq:int-AA}
\end{equation}
which, in terms of the mode functions of \cref{EQ:FieldOperator_SSE}, can be written as
\begin{align}
 \llangle \hat{A}_{\mu}( q) \hat{A}_{\nu}(-q)\rrangle = \frac{VT}{(2\pi)^4}  \sum_{\sigma}  A_{\bq,\sigma} A^*_{\bq,\sigma} \, e^\sigma_\mu(\hat \bq) e^{\sigma \dag}_\nu(\hat \bq) \, \delta(q^0-\omega_{\bq,\sigma}),
 \label{eq:int-AAA}
\end{align}
where we identify the spacetime integral with the four-volume, i.e.
\begin{align}
(2\pi)^4\delta^4(0) = \int d^4x =VT.
\label{eq:app-delta0}
\end{align}

In this part, we explicitly compute the four distinct momentum integrals arising in \cref{eq:W1-int}. We begin with the first integral
\begin{align}
I_1^{\mu\nu} = &\int dp_1^4 \int dp_2^4  \,  \delta^{(4)}(q-p_1+p_2)  q^{\nu} q^{\mu}\, \delta(p_1^2+m^2) \delta(p_2^2+m^2) \Theta(p_1^0) \Theta(-p_2^0) \nonumber\\
& = 2\pi \, q^\mu q^\nu \, \left(1+\frac{4m^2}{q^2}\right)^{\frac12} \, \Theta\left(1+\frac{4m^2}{q^2}\right).
\end{align}
The second integral is
\begin{align}
I_2^{\mu\nu} = &\int d^4p_1 \int d^4p_2  \,  \delta^{(4)}(q-p_1+p_2)  p_{1}^{\nu} p_2^{\mu}\, \delta(p_1^2+m^2) \delta(p_2^2+m^2) \Theta(p_1^0) \Theta(-p_2^0) \nonumber\\
& = \pi \left(1+\frac{4m^2}{q^2}\right)^{\frac12} \, \Theta\left(1+\frac{4m^2}{q^2}\right) \left[ \mathcal{A} \, \frac{q^\mu q^\nu}{q^2} + \frac{1}{3} (\mathcal{B}- \mathcal{A})\left(\eta^{\mu\nu} - \frac{q^\mu q^\nu}{q^2}\right)\right],
\end{align}
where $\mathcal{A}$ can be determined as
\begin{align}
\frac{q_{\mu} q_{\nu}}{q^2}I_2^{\mu\nu} & = \pi \left(1+\frac{4m^2}{q^2}\right)^{\frac12} \Theta(1+\frac{4m^2}{q^2}) \, \mathcal{A} \nonumber\\
& = -\frac{\pi}{2} q^2 \left(1+\frac{4m^2}{q^2}\right)^{\frac12} \Theta\left(1+\frac{4m^2}{q^2}\right),
\end{align}
and $\mathcal{B}$ is given as
\begin{align}
\eta_{\mu\nu} I_2^{\mu\nu} & = \pi \left(1+\frac{4m^2}{q^2}\right)^{\frac12}  \Theta\left(1+\frac{4m^2}{q^2}\right) \,  \mathcal{B} \nonumber\\
& = -\pi (q^2+2m^2) \left(1+\frac{4m^2}{q^2}\right)^{\frac12}  \Theta\left(1+\frac{4m^2}{q^2}\right).
\end{align}
From the above, we find the explicit form of $I_2^{\mu\nu}$ as
\begin{align}
I_2^{\mu\nu}  = \frac{2\pi}{12} \left(1+\frac{4m^2}{q^2}\right)^{\frac12} \, \Theta\left(1+\frac{4m^2}{q^2}\right) \left[ - 3 q^\mu q^\nu - (4m^2+q^2)\left(\eta^{\mu\nu} - \frac{q^\mu q^\nu}{q^2}\right)\right].
\end{align}
The third momentum integral is
\begin{align}
I_3^{\mu\nu} = &\int d^4p_1 \int d^4p_2  \,  \delta^{(4)}(q-p_1+p_2)  p_{1}^{\nu} q^{\mu}\, \delta(p_1^2+m^2) \delta(p_2^2+m^2) \Theta(p_1^0) \Theta(-p_2^0) \nonumber\\
& = \pi \left(1+\frac{4m^2}{q^2}\right)^{\frac12} \, \Theta\left(1+\frac{4m^2}{q^2}\right)  q^\mu q^\nu.
\end{align}
Finally, the last integral is
\begin{align}
I_4^{\mu\nu} = & \int d^4p_1 \int d^4p_2  \,  \delta^{(4)}(q-p_1+p_2)  p_{2}^{\nu} q^{\mu}\, \delta(p_1^2+m^2) \delta(p_2^2+m^2) \Theta(p_1^0) \Theta(-p_2^0) \nonumber\\
& = - I_3^{\mu\nu}.
\end{align}
From the combination of the above, we find 
\begin{align}
& \int\!\!\!\int_{p_1,p_2}  (2p_{1}-q)^{\nu} (2p_2+q)^{\mu}\,\delta^{(4)}(p_1^2+m^2)  \delta^{(4)}(q-p_1+p_2)  \delta^{(4)}(p_2^2+m^2)  \, \Theta(p_1^0) \Theta(-p_2^0) \nonumber\\
 & =    (-I_1+4I_2+4I_3)^{\mu\nu}= \frac{2\pi}{3} (-q^2) \left(1+\frac{4m^2}{q^2}\right)^{\frac32} \, \Theta\left(1+\frac{4m^2}{q^2}\right) \left(\eta^{\mu\nu} - \frac{q^\mu q^\nu}{q^2}\right).
 \label{eq:app-momentum}
\end{align}
Now we turn to the gauge-invariant form of the effective action in momentum space. We can write the gauge field combination in terms of the electromagnetic field tensor as
\begin{align}
  - (q^2 \, \eta^{\mu\nu} - q^\mu q^\nu) \,  \llangle \hat{A}_{\mu}(q) \hat{A}_{\nu}(-q)\rrangle = -\frac12 \llangle \hat{F}_{\mu\nu}(q) \hat{F}^{\mu\nu}(-q) \rrangle.
  \label{eq:app-ff}
\end{align}
The explicit form of the quadratic invariant reads
\begin{align}
\int d^4x \,\llangle - F_{\mu\nu}(x) F^{\mu\nu}(x) \rrangle
= VT \sum_{\sigma} \int_q \,
  \llangle - F^{\mu\nu}_{\sigma}(q)\,F^{\mu\nu}_{\sigma}(-q)\rrangle,
  \label{eq:app-fff}   
\end{align}
where for real fields one has $F_{\mu\nu}(-q) = F^*_{\mu\nu}(q)$.
Substituting the above expression into \cref{eq:W1-int}, we arrive at \cref{eq:Gamma-Boson}.  Similarly, the leading-order pair-production probability for charged fermions reads
\begin{align}
  \mathcal{P}^{f}_{\text{decay}}
  = \frac{4g^2\pi}{3} \int_q 
  \Theta\!\left(-q^2 - 4m^2\right)
  \left(1 - \frac{2m^2}{q^2}\right)
  \left(1 + \frac{4m^2}{q^2}\right)^{\!1/2}
  \llangle -\hat{F}_{\mu\nu}(q)\, \hat{F}^{\mu\nu}(-q) \rrangle.
 \label{eq:n-omega-fermion}
\end{align}


\section{Complete Analytical Expressions}\label{sec:app-compute}

In the main text of \cref{Sec:Phenomenology_I} and \cref{Sec:Phenomenology_II}, we reported the final expressions of our results in the physically relevant parameter regimes, where the formulas take a simplified form. For completeness, in this appendix, we present the full exact analytical solutions, valid for general parameter values. These results are expressed in terms of special functions and provide the basis from which the approximate formulas were derived.

\subsection*{C.1. ~ High-Energy Astrophysics}\label{sec-app-HEA}
Starting from \cref{eq:n-omega-EM} and \cref{eq:n-omega-A}, we can compute the vacuum decay rate to a scalar species in the presence of a macroscopic charged background as 
\begin{align}
\Gamma^b = \frac{1}{T}\int d\omega \, \llangle n^b({\omega}) \rrangle\, \propto\, \frac{g^2Q^2}{3\pi^2}
\, \left(1 - \frac{4m^2}{\omega_\star^2 b^2}\right)^{3/2}
\Theta\!\left(\omega_\star b -  2m\right) \mathcal{D},
  \label{eq:n-omega-}
\end{align} 
where $y=\frac{\omega}{\omega_\star}$ and $\mathcal{D}$ is defined as
\begin{align}
\begin{split}
    \mathcal{D}(b;\,n_1,\,n_2) &\equiv \int_{\frac{\Lambda_\text{IR}}{\omega_\star}}^1 \, dy \, y^{n_1} \left( y^2 - b^2\right)^{\frac12} + \int^{\frac{\Lambda_\text{UV}}{\omega_\star}}_1 \, dy \, y^{n_2} \left( y^2 - b^2\right)^{\frac12}\\
    &=\mathcal{D}_{\text{IR}}(b;\,n_1)+\mathcal{D}_{\text{UV}}(b;\,n_2),
\end{split}
\end{align}
in which the parameters $n_{1,2}$ and $b$ take the following values for the electromagnetic (EM) and dark-photon cases:
\begin{align}
    n_1 &= -1 + \gamma_T, & n_2 &= 1 - \delta_T, & b &= \frac{\omega_p}{\omega_\star}, && \text{(EM)},\\[4pt]
    n_1 &= -3 + \gamma_L, & n_2 &= -1 - \delta_L, & b &= \frac{m_{A'}}{\omega_\star}, && \text{(dark photon)}.
\end{align}
Making use of the changes of variable $y^2\mapsto t + b^2$, the integrals above can be expressed in terms of the hypergeometric representation introduced in \cref{eq:app-Hyp---} as
\begin{equation}
\int dy\, y^{n}\,\sqrt{y^{2}-b^{2}}=\frac{b^{\,n-1}}{3}\,(y^2-b^2)^{3/2}\,
{}_2F_1\!\left(\frac{3}{2},\,\frac{1-n}{2};\,\frac{5}{2};\,1-\frac{y^2}{b^{2}}\right)
+ C.
\end{equation}
Solutions for $\mathcal{D}_\text{EM}$ and $\mathcal{D}_{A'}$ are given by
\begin{align}
\begin{split}
\mathcal{D}_\text{EM} &= \frac{1}{3}\, \left[ \left(\frac{\omega_p}{\omega_\star}\right)^{\gamma_T-2} \left(y^2-\frac{\omega_p^2}{\omega_\star^2}\right)^{3/2}\,
\left.{}_2F_1\!\left(\frac{3}{2},\,\frac{2-\gamma_T}{2};\,\frac{5}{2};\,1-\frac{\omega_\star^2y^2}{\omega_p^2}\right)\right|_{y={\Lambda_\text{IR}}/{\omega_\star}}^{y=1} \right.\\
 &\quad \left. +
\left(\frac{\omega_p}{\omega_\star}\right)^{-\delta_T} \left(y^2-\frac{\omega_p^2}{\omega_\star^2}\right)^{3/2}\,
\left.{}_2F_1\!\left(\frac{3}{2},\frac{\delta_T}{2};\,\frac{5}{2};\,1-\frac{\omega_\star^2y^2}{\omega_p^2}\right)\right|^{y={\Lambda_\text{UV}}/{\omega_\star}}_{y=1}\right]
\label{eq:app-Hyp---D-gen}
\end{split}
\end{align} 
and
\begin{align}
\begin{split}
    \mathcal{D}_{A'} &= \frac{1}{3}\left[\left(\frac{m_{A'}}{\omega_\star}\right)^{\gamma_L-4}\left(y^2-\frac{m_{A'}^2}{\omega_\star^2}\right)^{3/2}\ \left. _2F_1\left(\frac{3}{2},\frac{4-\gamma_L}{2};\frac{5}{2};1-\frac{\omega_\star^2 y^2}{m_{A'}^2}\right)\right|_{y=\Lambda_{\text{IR}}/\omega_\star}^{y=1}\right.\\
    &\quad\left.+\left(\frac{m_{A'}}{\omega_\star}\right)^{-(\delta_L+2)}\left(y^2-\frac{m_{A'}^2}{\omega_\star^2}\right)^{3/2}\ \left. _2F_1\left(\frac{3}{2},\frac{\delta_L+2}{2};\frac{5}{2};1-\frac{\omega_\star^2 y^2}{m_{A'}^2}\right)\right|_{y=1}^{y=\Lambda_{\text{UV}}/\omega_\star}\right]. 
\end{split}
\end{align}
To simplify our results,  we consider the typical regime \(\tfrac{\Lambda_{\mathrm{IR}}}{\omega_p} \simeq 1\),
\(\tfrac{\Lambda_{\mathrm{UV}}}{\omega_p} \gg 1\), and $\omega_\star \gtrsim \omega_p$, which after making use of the limit in \cref{eq:2F1-0} become
\begin{align}
\begin{split}
\mathcal{D}_\text{EM} &\approx \frac{1}{3}\, \left[ \left(\frac{\omega_p}{\omega_\star}\right)^{\gamma_T-2} \left(1-\frac{\omega_p^2}{\omega_\star^2}\right)^{3/2}\,
{}_2F_1\!\left(\frac{3}{2},\,\frac{2-\gamma_T}{2};\,\frac{5}{2};\,1-\frac{\omega_\star^2}{\omega_p^2}\right)
\right]\\ &\quad+\mathcal{D}_{\text{UV}}\left(\frac{\omega_p}{\omega_\star};\, 1-\delta_T\right),
\label{eq:app-Hyp---D-EM}
\end{split}
1\end{align} 
and
\begin{align}
\begin{split}
\mathcal{D}_{A'} &\approx \frac{1}{3}\, \left[ \left(\frac{m_{A'}}{\omega_\star}\right)^{\gamma_L-4} \left(1-\frac{m_{A'}^2}{\omega_\star^2}\right)^{3/2}\,
{}_2F_1\!\left(\frac{3}{2},\,\frac{4-\gamma_L}{2};\,\frac{5}{2};\,1-\frac{\omega_\star^2}{m_{A'}^2}\right) \right]\\ &\quad+\mathcal{D}_{\text{UV}}\left(\frac{m_{A'}}{\omega_\star};\, -\delta_T-1\right),
\label{eq:app-Hyp---D-A}
\end{split}
\end{align} 
where $\mathcal{D}_{\text{UV}}$ takes the form
\begin{equation}
  \mathcal{D}_{\text{UV}}(b;\, n_2)   \approx    \begin{cases}
  \frac{b^{n_2-1}}{n_2+2}\left(\frac{\Lambda_{\text{UV}}}{\omega_\star b}\right)^{n_2+2}  & n_2>-2\\
  &\\
\ln\left( \frac{\Lambda_\text{UV}}{\omega_\star b}\right) & n_2 =-2 \\
 &\\
 \frac{b^{n_2+2}\sqrt{\pi}}{4} \, \frac{\Gamma\left(-\frac{n_2+2}{2}\right)}{\Gamma\left(\frac{1-n_2}{2}\right)} & n<-2
    \end{cases}.
\end{equation}

\subsection*{C.2 ~ Axion-QED Reheating}\label{sec-app-a-QED}

Here, we detail the integration steps deferred from \cref{Sec:Phenomenology_II} and provide a transparent derivation of the result. Starting from the above, we use \cref{eq:Gamma-Boson} to compute the rate of created boson pairs with mass $m$ as 

\begin{align}
  \mathcal{P}^b_{\chi} \approx   \frac{2.7 g^2 \,Q^2\,  V}{6\pi \, \xi^2}\,\int_H^{q_{\text{max}}} d|\bq||\bq|\int_{\sqrt{|\bq|^2+4m^2}}^{\infty} d\omega\  \left(1-\frac{4m^2}{\omega^2-|\bq|^2}\right)^{\frac{3}{2}}{\left(\frac{\omega^2}{|\bq|^2}-1\right)} \,  \, e^{- \tfrac{1}{\xi^2}\tfrac{\omega^2}{|\bq|^2}},
\end{align}
where we have approximated Euler's number as $e\approx 2.7$ and $q_{\text{max}}=m_\chi\xi$ as dictated by the instability window \cref{eq:inst}. As a result the number density of the generated pairs in \cref{eq:a-QED} can be written as
\begin{align}
    n^b_{\chi} = 2m \, \int_{{H}/{2m}}^{{q_{\text{max}}}/{2m}} \, {d z} \   n_{\chi}(z),
\end{align}
in which $z=|\bq|/2m$. This induces the following re-parametrisation on the spectral number density of bosons 
\begin{equation}
    n^b_{\chi}(z) = \frac{2.7 g^2 \,Q^2(2m)^2 }{6\pi \, \xi^2} \, \mathcal{I}_b(z), 
\label{eq:ab-QED---}
\end{equation}

\begin{figure}[H]
\centering
\begin{subfigure}{0.45\textwidth}
    \centering
    \includegraphics[width=\linewidth]{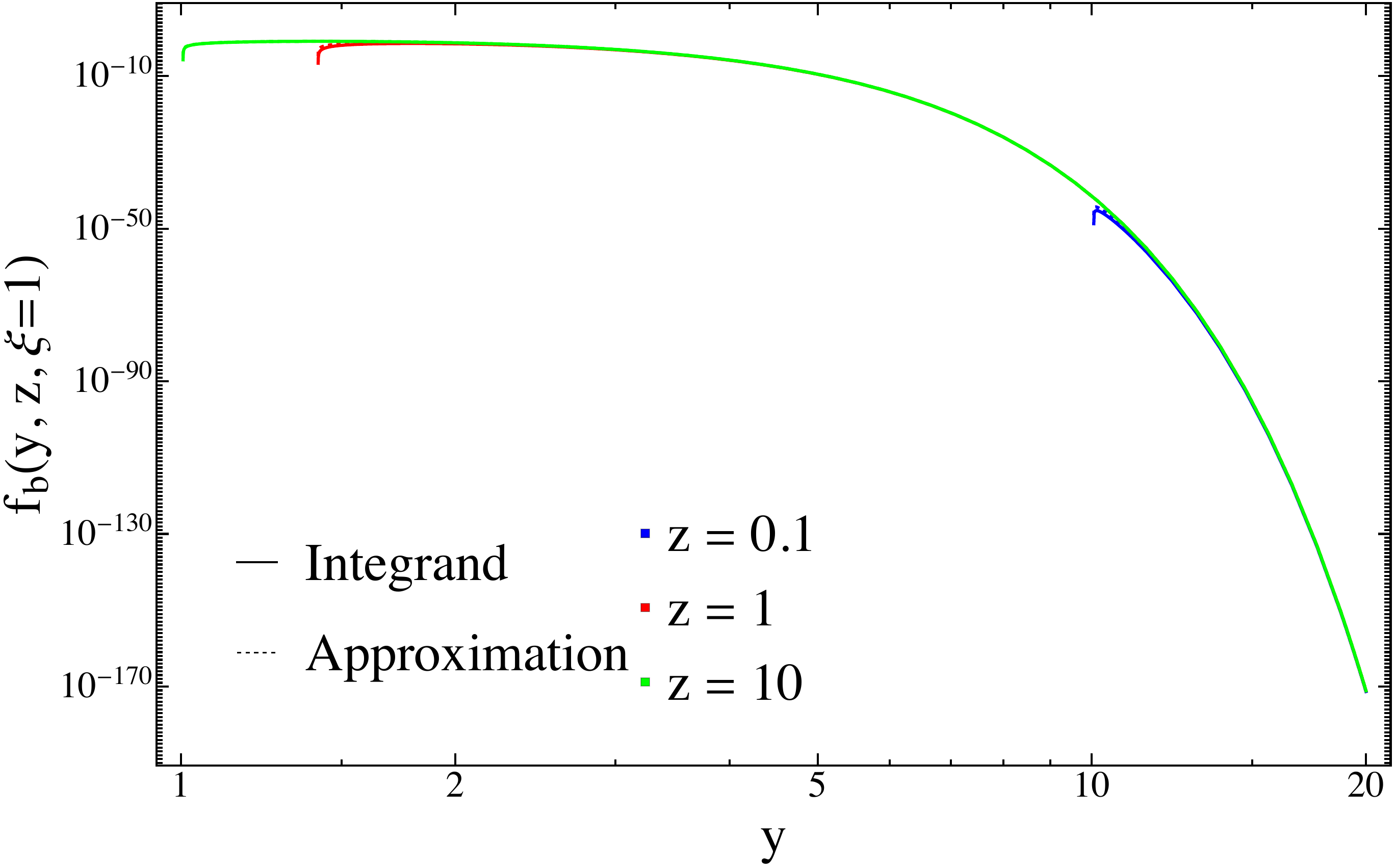}
\end{subfigure}
\hfill
\begin{subfigure}{0.45\textwidth}
    \centering
    \includegraphics[width=\linewidth]{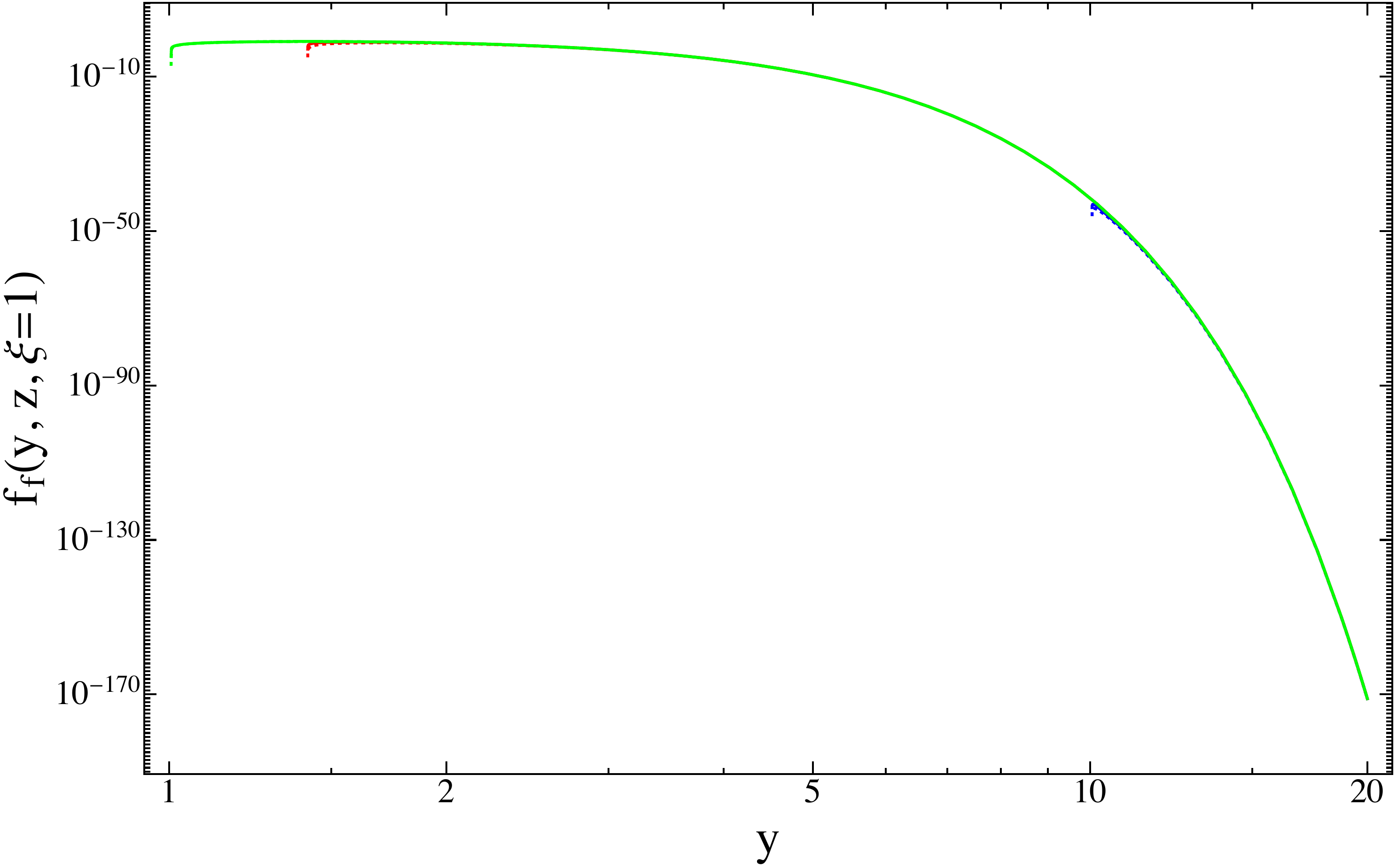}
\end{subfigure}
\begin{subfigure}{0.45\textwidth}
    \centering
    \includegraphics[width=\linewidth]{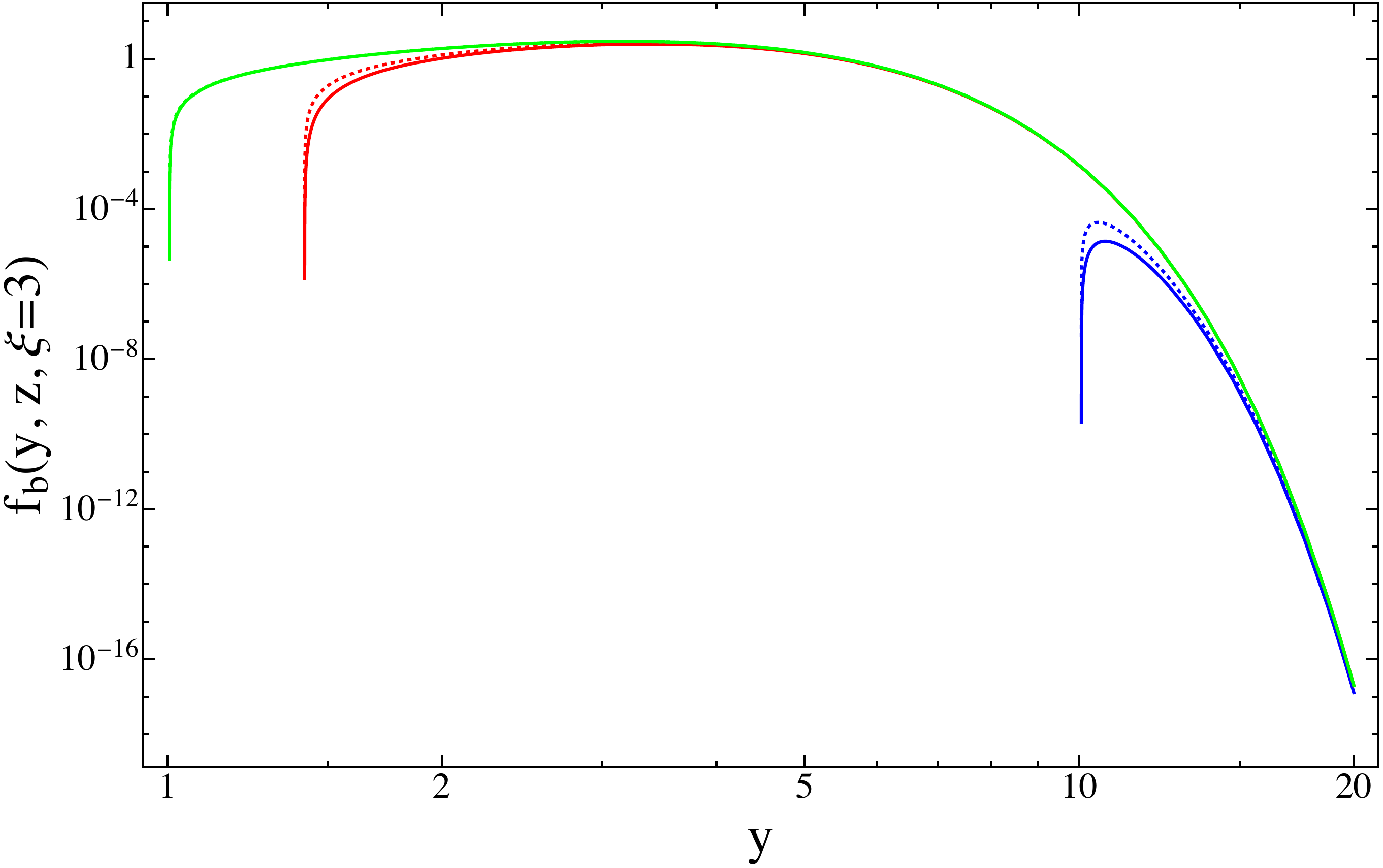}
\end{subfigure}
\hfill
\begin{subfigure}{0.45\textwidth}
    \centering
    \includegraphics[width=\linewidth]{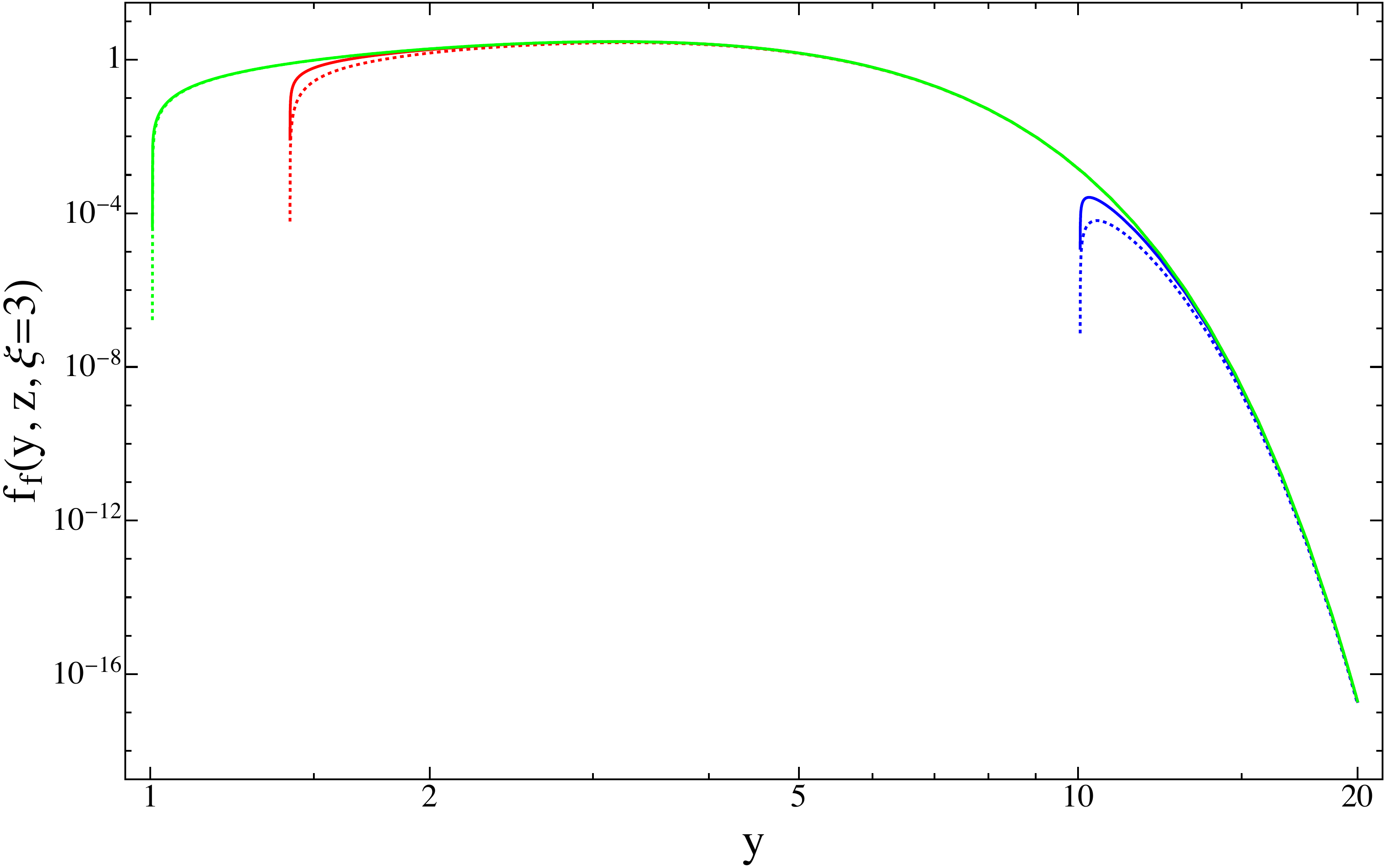}
\end{subfigure}
\begin{subfigure}{0.45\textwidth}
    \centering
    \includegraphics[width=\linewidth]{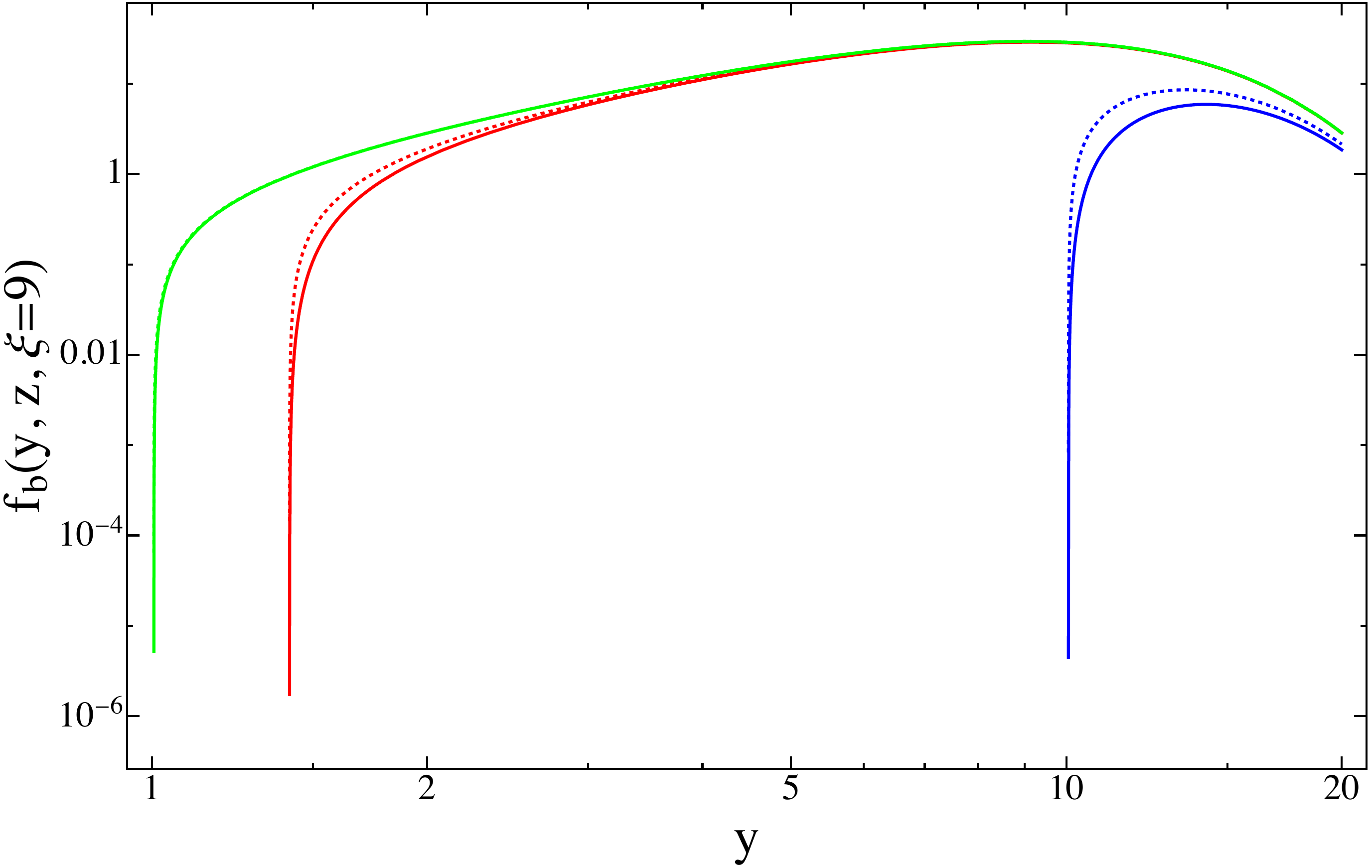}
\end{subfigure}
\hfill
\begin{subfigure}{0.45\textwidth}
    \centering
    \includegraphics[width=\linewidth]{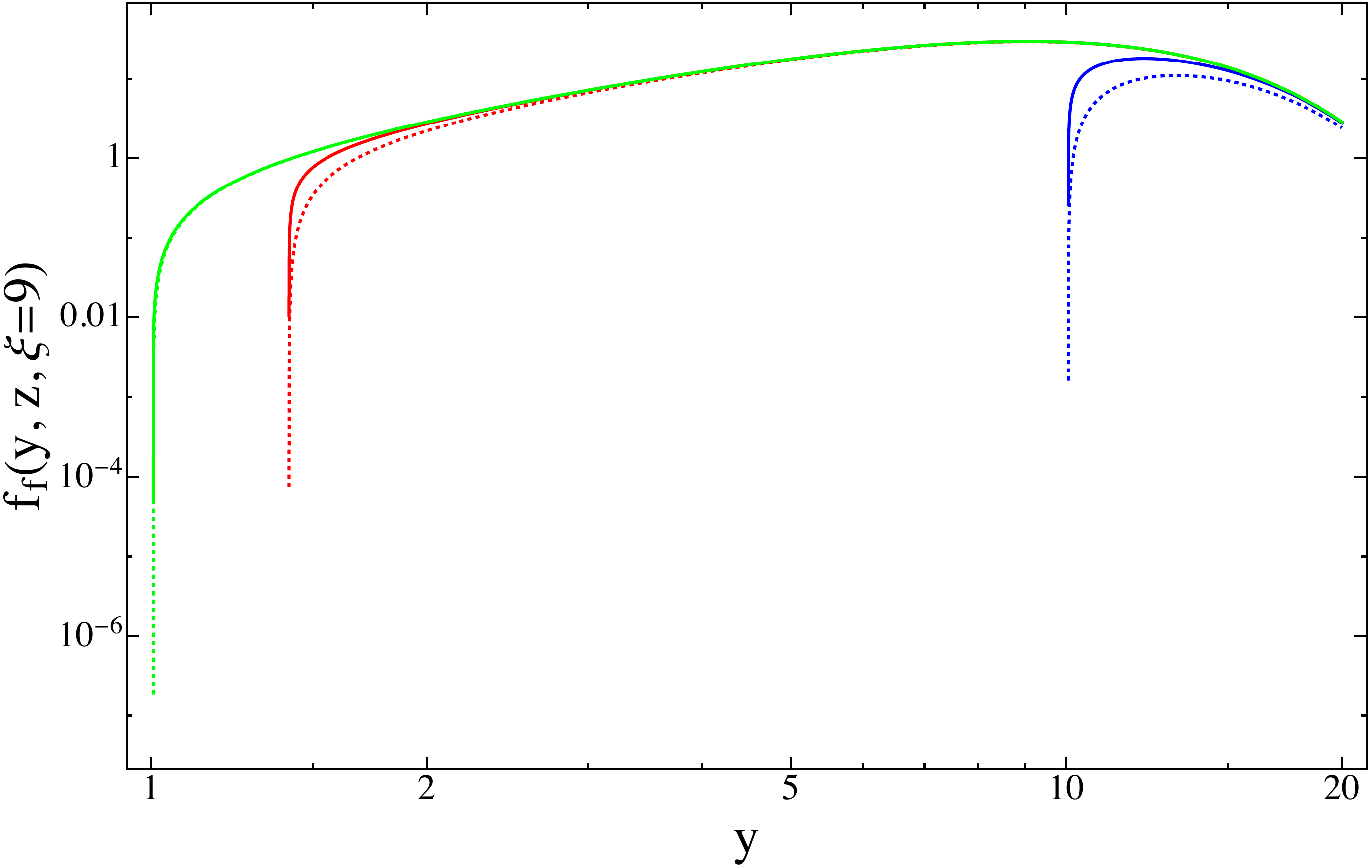}
\end{subfigure}
\begin{subfigure}{0.45\textwidth}
    \centering
    \includegraphics[width=\linewidth]{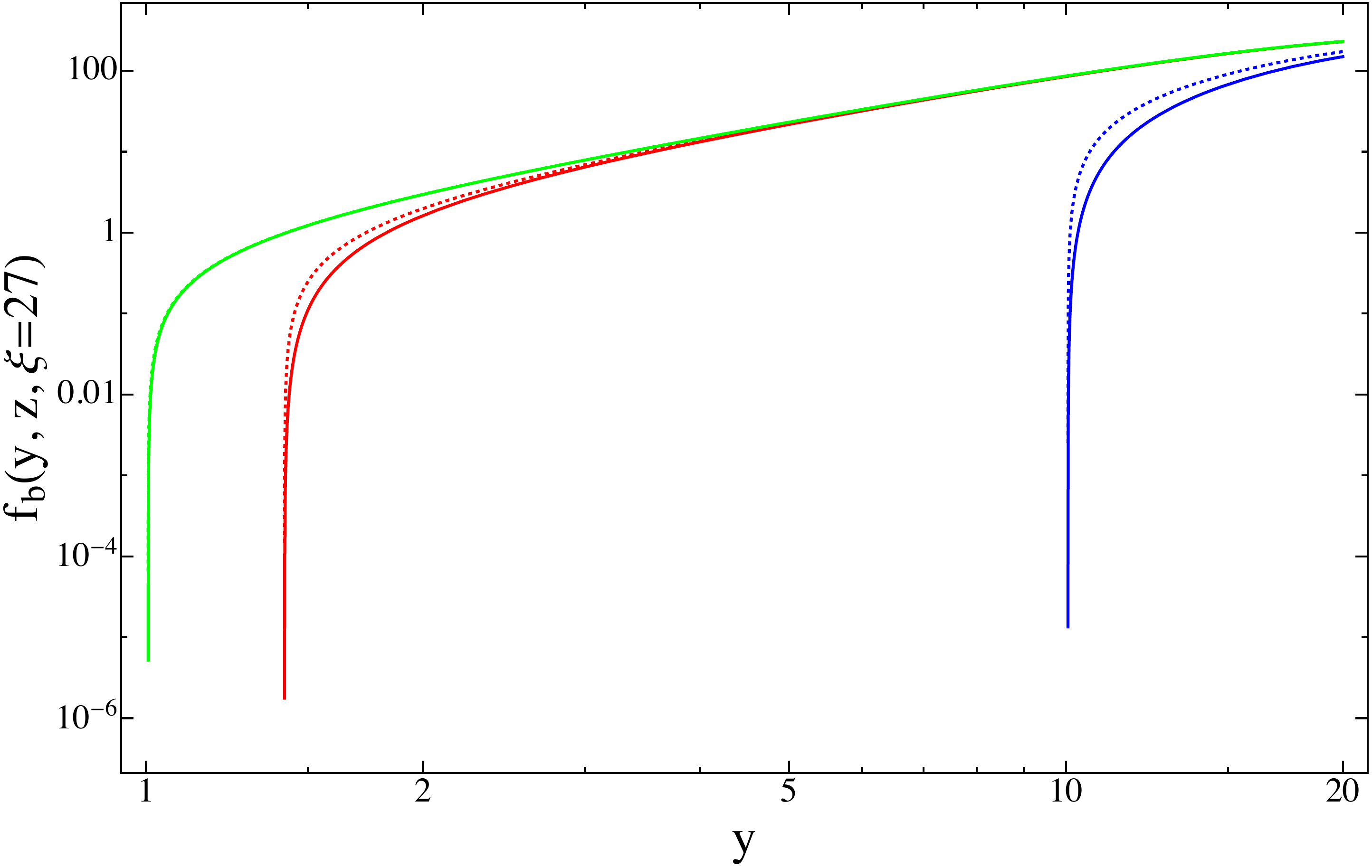}
\end{subfigure}
\hfill
\begin{subfigure}{0.45\textwidth}
    \centering
    \includegraphics[width=\linewidth]{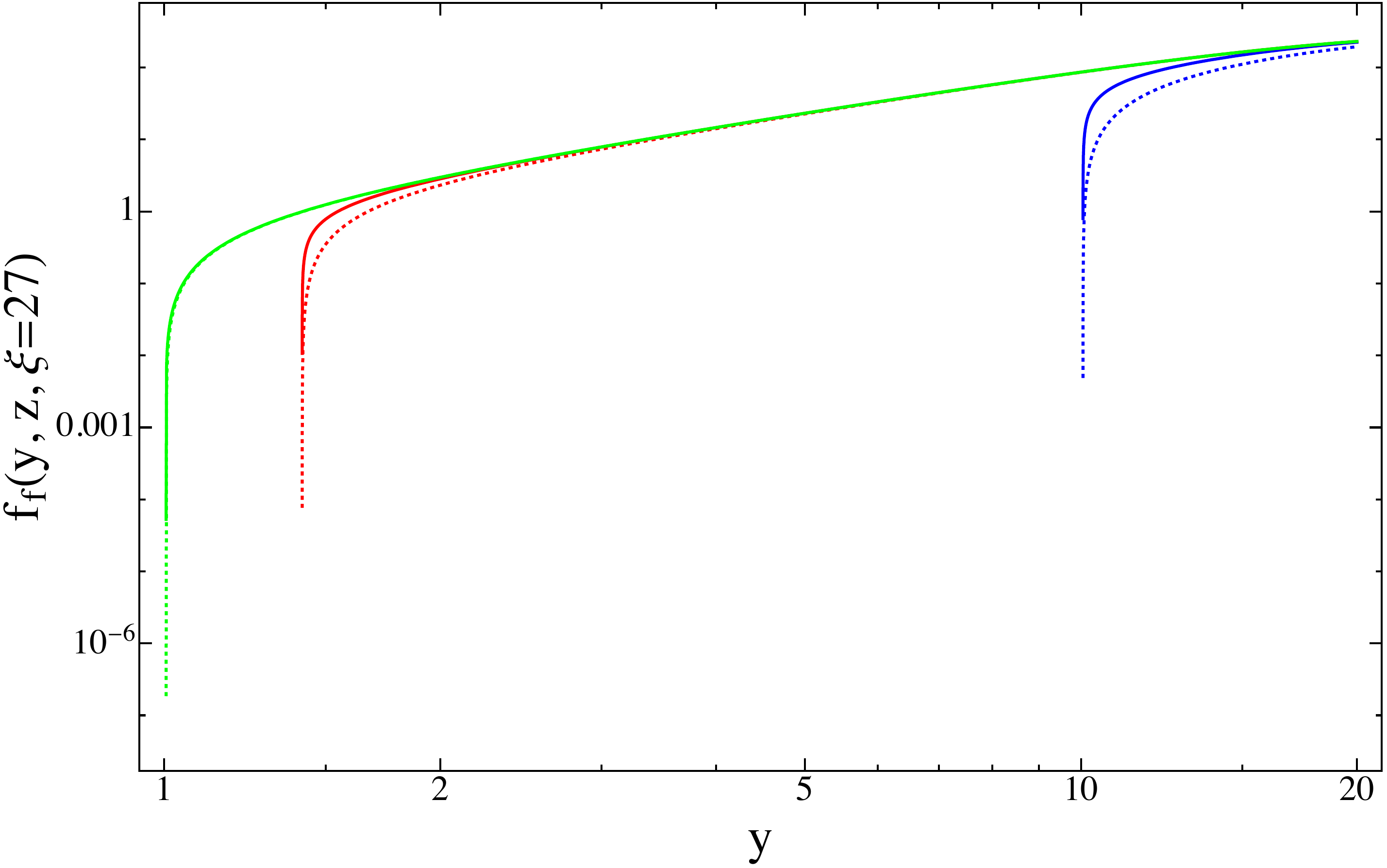}
\end{subfigure}

\caption{Comparison of the exact integrand (solid lines) with the analytical approximation (dashed lines) for different values of $z$ and $\xi$. Note that both integrands are defined over the domain $y \in [\sqrt{1+z^{-2}},\,\infty)$. 
The left panel shows the bosonic integrand in \cref{eq:app-fbb}, while the right panel shows the fermionic integrand in \cref{eq:app-f-f} in the right panels. The proposed approximation reproduces the exact integrand with excellent accuracy across different choices of $z$ and $\xi$ successfully capturing both the near-threshold behaviour and the peak structure. The quality of the fit improves with increasing $z$, precisely in the region that dominates the contribution to the resulting integral.
}
\label{fig:Boson-Fermion_approx}
\end{figure}

\noindent where we have performed the change of variables $\omega/|\bq |\mapsto y$ and $\mathcal{I}_b(z) $ is 
\begin{align}
 \mathcal{I}_b(z) \equiv   z^2 \, \int_{\sqrt{1+z^{-2}}}^{\infty} dy \, \left(1-\frac{z^{-2}}{y^2-1}\right)^{\frac{3}{2}}{(y^2-1)} \,  \, e^{ - y^2/\xi^2}.
 \label{eq:app-I-b}
\end{align}
Although the above integral lacks a closed-form solution, as in the bosonic case, an approximate form of the integrand can be devised to enable analytic evaluation
\begin{align}
\begin{split}
 f_b(y,z,\xi) & \equiv \left(1-\frac{z^{-2}}{y^2-1}\right)^{\frac{3}{2}}{(y^2-1)} \,  \, e^{ - y^2/\xi^2}  \approx    \left(y^2- 1 - z^{-2}\right)  \, e^{ - y^2/\xi^2}.
\label{eq:app-fbb}
\end{split}
\end{align}

Note that a direct expansion of the integrand about the threshold $y_0=\sqrt{1+z^{-2}}$ yields a local (Puiseux) series that captures the cusp, but deteriorates near the peak of the Gaussian–weighted profile. The factor $e^{-y^{2}/\xi^2}$ pushes the maximum to $y_{\rm peak}\gtrsim y_0$, typically outside the range where a low-order expansion around $y_0$ remains accurate. By contrast, our threshold-matched ansatz enforces the exact zero at $y_0$ and preserves the correct near-threshold curvature while remaining accurate through the peak, thus providing a uniformly good fit over the region that dominates the integral, see left panel of Fig \ref{fig:Boson-Fermion_approx}.

Using the ansatz \cref{eq:app-fbb} in \cref{eq:app-I-b}, we can write the integral as
\begin{align}
\begin{split}
        \mathcal{I}_b(z) &\approx  z^2 \, \int_{\sqrt{1+z^{-2}}}^{\infty} dy \, \left(y^2- 1 - z^{-2}\right)\,  \, e^{ - y^2/\xi^2}\\
        &=\frac{z^2\xi^2}{2}  \left[  \sqrt{1+z^{-2}} e^{-\frac{(1+z^{-2})}{\xi ^2}}+\frac{\sqrt{\pi }}{2\xi} \left(\xi^2-{2}(1+{z^{-2}})\right) \text{erfc}\left(\frac{\sqrt{1+z^{-2}}}{\xi }\right)\right],
\label{eq:app-reheating-b}
\end{split}
\end{align}
where $\text{erfc}(z)$ is the complementary error function defined in \cref{eq:app-error-c}. We can approximate the function above by studying its asymptotic behaviour relative to the combination $\xi z=1$, 
\begin{equation}
\mathcal{I}_b(z)\approx\begin{cases}
       \frac{z^2\xi^2}{2}  \left[ e^{-\frac{1}{\xi ^2}}+\frac{\sqrt{\pi }}{2\xi} \left(\xi ^2-2\right) \text{erfc}\left(\frac{1}{\xi }\right)\right] & \xi z\gtrsim1,\\
       \\
       \frac{z^3\xi^4}{2}e^{-\frac{1+z^{-2}}{\xi^2}}& \xi z\lesssim1.
    \end{cases}
    \label{eq:app-I-b-z}
\end{equation}
The spectral number density can then be written as 
\begin{align}
\begin{split}
    n^b_\chi(|\bq|)&\approx\frac{2.7g^2Q^2}{12\pi}|\bq|^2\left[\left( e^{-\frac{1}{\xi ^2}}+\frac{\sqrt{\pi }}{2\xi} \left(\xi ^2-2\right) \text{erfc}\left(\frac{1}{\xi }\right)\right)\Theta(\xi |\bq|-2m)\right.\\
    &\quad+\left.\frac{|\bq|}{2m}\xi^2\exp\left(-\frac{1}{\xi^2 }\left(1+\frac{4m^2}{|\bq|^2}\right)\right)\Theta(2m-\xi|\bq|)\right]
\end{split}
\label{eq:spectraltotal}
\end{align}

From \cref{eq:spectraltotal}, we see that for $\xi |\bq|\lesssim 2m$, the  spectral number density remains negligible, as it is heavily suppressed by the exponential factor. This allows us to approximate the total number density as
\begin{equation}
    n^b_{\chi} \approx \frac{2.7 g^2 \,Q^2 }{36\pi} \, \frac{\sqrt{\pi}}{2}\, m_{\chi}^3 \, \xi^4 \, \mathcal{D}_{\chi}\Theta(m_\chi\xi^2-2m),
\label{eq:ab-QED---o-}
\end{equation}
in which we used $|\bq_\text{max}|=m_\chi \xi$, and $\mathcal{D}_\chi$ is 
\begin{equation}
    \mathcal{D}_{\chi}\equiv\left[ \frac{2}{\sqrt{\pi}\xi} e^{-\frac{1}{\xi ^2}}+\frac{1}{\xi^2} \left(\xi ^2-2\right) \text{erfc}\left(\frac{1}{\xi }\right)\right].
\end{equation}
The parameter $ \mathcal{D}_{\chi} \in (0.4, 1)$ is defined such that $ \mathcal{D}_{\chi} = 0.4$ at $\xi = 1$ and increases monotonically, asymptotically approaching unity in the limit of large $\xi$. Finally, we find the background decay rate to bosons as
\begin{equation}  \Gamma_{\chi}=\int_{H}^{q_\text{max}} d|\bq|\ \frac{n_{\chi}(|\bq|)}{\Delta t{(|\bq|)}} \approx \xi \int_{H}^{q_\text{max}} d|\bq|\ |\bq|\,{n_\chi(|\bq|)}.
\label{eq:app-Gamma-chi-QED-b}
\end{equation}
which gives 
\begin{equation}
   \Gamma^b_{\chi} \approx \frac{2.7 g^2 \,Q^2 }{12\pi} \, \frac{\sqrt{\pi}}{8}\, m_{\chi}^4 \, \xi^5 \, \mathcal{D}_{\chi} \Theta(m_\chi\xi^2-2m).
\end{equation}

Now we turn to the fermionic case. In complete analogy, we introduce the following reparametrisation for the spectral number density of fermions
\begin{equation}
    n^f_{\chi}(z) = \frac{5.4 g^2 \,Q^2(2m)^2 }{3\pi \, \xi^2} \, \mathcal{I}_f(z), 
\label{eq:af-QED---}
\end{equation}
where 
\begin{align}
 \mathcal{I}_f(z) \equiv   z^2 \, \int_{\sqrt{1+z^{-2}}}^{\infty} dy \, \left(1+\frac{1}{2}\frac{z^{-2}}{y^2-1}\right)\left(1-\frac{z^{-2}}{y^2-1}\right)^{\frac{1}{2}}{(y^2-1)} \,  \, e^{ - y^2/\xi^2}.
\end{align}
While the integral cannot be solved analytically, similar to the bosonic case, the integrand can be approximated in a way that renders the expression analytically tractable
\begin{align}
\begin{split}
 f_f(y,z,\xi) & \equiv \left(1+\frac{1}{2}\frac{z^{-2}}{y^2-1}\right)\left(1-\frac{z^{-2}}{y^2-1}\right)^{\frac{1}{2}}{(y^2-1)} \,  \, e^{ - y^2/\xi^2} \\
& \approx    \left(y^2- 1 - \frac{1}{2}z^{-2}-\frac{1}{2}\frac{z^{-4}}{y^2}\right)  \, e^{ - y^2/\xi^2}.
\label{eq:app-f-f}
\end{split}
\end{align}
The right panels of \cref{fig:Boson-Fermion_approx} display a comparison between the exact integrand and our approximation, demonstrating very good agreement.
With the above approximation, the integral admits an analytic solution expressed in terms of the error function
\begin{align}
    \begin{split}
        \mathcal{I}_f(z)& \approx  z^2 \, \int_{\sqrt{1+z^{-2}}}^{\infty} dy\, \left(y^2- 1 - \frac{1}{2}z^{-2}-\frac{1}{2}\frac{z^{-4}}{y^2}\right)  \, e^{ - y^2/\xi^2}\\
        &=\frac{z^2}{2}\left[\frac{\left(\xi^2(1+z^{-2})-z^{-4}\right)}{\sqrt{1+z^{-2}}}e^{-\frac{(1+z^{-2})}{\xi^2}}+\frac{\sqrt{\pi}}{2\xi}\left(\xi^2(\xi^2-2-z^{-2})+2z^{-4}\right)\text{erfc}\left(\frac{\sqrt{1+z^{-2}}}{\xi}\right)\right]
\label{eq:app-reheating-f}
\end{split}
\end{align}
The function takes different analytic forms in the regimes 
$\xi z \gtrsim 1$ and $\xi z \lesssim 1$, given respectively by
\begin{equation}
    \mathcal{I}_f(z)\approx\begin{cases}
       \frac{z^2\xi^2}{2}\left[e^{-\frac{1}{\xi^2}}+\frac{\sqrt{\pi}}{2\xi}(\xi^2-2)\text{erfc}\left(\frac{1}{\xi}\right)\right] & \xi z\gtrsim1,\\
       \\
       \frac{3z^3\xi^4}{4}e^{-\frac{(1+z^{-2})}{\xi^2}}& \xi z\lesssim1.
    \end{cases}
\end{equation}
This is in turn allows us to write the spectral density as
\begin{align}
\begin{split}
    n^f_\chi(|\bq|)&\approx\frac{5.4g^2Q^2}{6\pi}|\bq|^2\left[\left( e^{-\frac{1}{\xi ^2}}+\frac{\sqrt{\pi }}{2\xi} \left(\xi ^2-2\right) \text{erfc}\left(\frac{1}{\xi }\right)\right)\Theta(\xi |\bq|-2m)\right.\\
    &\quad+\left.\frac{3|\bq|}{2m}\xi^2\exp\left(-\frac{1}{\xi^2 }\left(1+\frac{4m^2}{|\bq|^2}\right)\right)\Theta(2m-\xi|\bq|)\right]
\end{split}
\label{eq:spectraltotalf}
\end{align}
This implies that, similar to the bosonic case, the above expression remains negligible for 
$\xi |\bq| \lesssim 2m$ and becomes significant once $\xi |\bq| \gtrsim 2m$.
As a result, we find that total number density is given by
\begin{equation}
    n^f_{\chi} \approx \frac{5.4 g^2 \,Q^2 }{18\pi} \, \frac{\sqrt{\pi}}{2}\, m_{\chi}^3 \, \xi^4 \, \mathcal{D}_{\chi} \Theta(m_\chi\xi^2-2m).
\label{eq:ab-QED---o-g}
\end{equation}
Accordingly, the background decay rate into fermions reads
\begin{equation}
   \Gamma^f_{\chi} \approx \frac{5.7 g^2 \,Q^2 }{24\pi} \, \frac{\sqrt{\pi}}{2}\, m_{\chi}^4 \, \xi^5 \, \mathcal{D}_{\chi} \Theta(m_\chi\xi^2-2m).
\label{eq:ab-QED---o-decay-ffff}
\end{equation}

\providecommand{\href}[2]{#2}\begingroup\raggedright\endgroup


\begin{thebibliography}{10}



\bibitem{heisenberg1929folgerungen}
W.~Heisenberg and H.~Euler, \emph{Folgerungen aus der diracschen theorie des positrons}, \href{https://doi.org/10.1007/BF01339504}{\emph{Zeitschrift für Physik} {\bfseries 69} (1929) 742}.

\bibitem{RefWorks:sauter1931ber}
F.~Sauter, \emph{Über das verhalten eines elektrons im homogenen elektrischen feld nach der relativistischen theorie diracs}, \href{https://doi.org/10.1007/BF01339461}{\emph{The European physical journal. A, Hadrons and nuclei} {\bfseries 69} (1931) 742}.

\bibitem{Schwinger1951}
J.~Schwinger, \emph{On gauge invariance and vacuum polarization}, \href{https://doi.org/10.1103/PhysRev.82.664}{\emph{Physical Review} {\bfseries 82} (1951) 664}.

\bibitem{Berdyugin2022}
A.~I. Berdyugin, N.~Xin, H.~Gao, S.~Slizovskiy, Z.~Dong, S.~Bhattacharjee et~al., \emph{Out-of-equilibrium criticalities in graphene superlattices}, \href{https://doi.org/10.1126/science.abi8627}{\emph{Science} {\bfseries 375} (2022) 430} [\href{https://arxiv.org/abs/https://www.science.org/doi/pdf/10.1126/science.abi8627}{{\ttfamily https://www.science.org/doi/pdf/10.1126/science.abi8627}}].

\bibitem{RefWorks:piñeiro2019sauterschwinger}
A.~M. Piñeiro, D.~Genkina, M.~Lu and I.~B. Spielman, \emph{Sauter-schwinger effect with a quantum gas}, \href{https://doi.org/10.1088/1367-2630/ab3840}{\emph{New journal of physics} {\bfseries 21} (2019) 83035}.

\bibitem{RefWorks:martinez2016realtime}
E.~A. Martinez, C.~A. Muschik, P.~Schindler, D.~Nigg, A.~Erhard, M.~Heyl et~al., \emph{Real-time dynamics of lattice gauge theories with a few-qubit quantum computer}, \href{https://doi.org/10.1038/nature18318}{\emph{Nature (London)} {\bfseries 534} (2016) 516}.

\bibitem{Hebenstreit:2011pm}
F.~Hebenstreit, \emph{{Schwinger effect in inhomogeneous electric fields}}, Ph.D. thesis, Graz U., 2011.
\newblock \href{https://arxiv.org/abs/1106.5965}{{\ttfamily 1106.5965}}.

\bibitem{Brezin:1970xf}
E.~Brezin and C.~Itzykson, \emph{{Pair production in vacuum by an alternating field}}, \href{https://doi.org/10.1103/PhysRevD.2.1191}{\emph{Phys. Rev. D} {\bfseries 2} (1970) 1191}.

\bibitem{Kim:2007pm}
S.~P. Kim and D.~N. Page, \emph{{Improved Approximations for Fermion Pair Production in Inhomogeneous Electric Fields}}, \href{https://doi.org/10.1103/PhysRevD.75.045013}{\emph{Phys. Rev. D} {\bfseries 75} (2007) 045013} [\href{https://arxiv.org/abs/hep-th/0701047}{{\ttfamily hep-th/0701047}}].

\bibitem{Hallin:1994ad}
J.~Hallin and P.~Liljenberg, \emph{{Fermionic and bosonic pair creation in an external electric field at finite temperature using the functional Schrodinger representation}}, \href{https://doi.org/10.1103/PhysRevD.52.1150}{\emph{Phys. Rev. D} {\bfseries 52} (1995) 1150} [\href{https://arxiv.org/abs/hep-th/9412188}{{\ttfamily hep-th/9412188}}].

\bibitem{Gould:2017fve}
O.~Gould and A.~Rajantie, \emph{{Thermal Schwinger pair production at arbitrary coupling}}, \href{https://doi.org/10.1103/PhysRevD.96.076002}{\emph{Phys. Rev. D} {\bfseries 96} (2017) 076002} [\href{https://arxiv.org/abs/1704.04801}{{\ttfamily 1704.04801}}].

\bibitem{Breit:1934zz}
G.~Breit and J.~A. Wheeler, \emph{{Collision of two light quanta}}, \href{https://doi.org/10.1103/PhysRev.46.1087}{\emph{Phys. Rev.} {\bfseries 46} (1934) 1087}.

\bibitem{STAR:2021twy}
{\scshape STAR} collaboration, J.~Adam et~al., \emph{{Observation of the Breit--Wheeler process and e$^+$e$^-$ pair production from real photons}}, \href{https://doi.org/10.1038/s41586-022-04565-8}{\emph{Nature} {\bfseries 606} (2022) 41} [\href{https://arxiv.org/abs/2109.00113}{{\ttfamily 2109.00113}}].

\bibitem{Lozanov:2018kpk}
K.~D. Lozanov, A.~Maleknejad and E.~Komatsu, \emph{{Schwinger Effect by an $SU(2)$ Gauge Field during Inflation}}, \href{https://doi.org/10.1007/JHEP02(2019)041}{\emph{JHEP} {\bfseries 02} (2019) 041} [\href{https://arxiv.org/abs/1805.09318}{{\ttfamily 1805.09318}}].

\bibitem{Mirzagholi:2019jeb}
L.~Mirzagholi, A.~Maleknejad and K.~D. Lozanov, \emph{{Production and backreaction of fermions from axion-$SU(2)$ gauge fields during inflation}}, \href{https://doi.org/10.1103/PhysRevD.101.083528}{\emph{Phys. Rev. D} {\bfseries 101} (2020) 083528} [\href{https://arxiv.org/abs/1905.09258}{{\ttfamily 1905.09258}}].

\bibitem{Maleknejad:2019hdr}
A.~Maleknejad, \emph{{Dark Fermions and Spontaneous $CP$ violation in $SU(2)$-axion Inflation}}, \href{https://doi.org/10.1007/JHEP07(2020)154}{\emph{JHEP} {\bfseries 07} (2020) 154} [\href{https://arxiv.org/abs/1909.11545}{{\ttfamily 1909.11545}}].

\bibitem{Domcke:2021fee}
V.~Domcke, Y.~Ema and K.~Mukaida, \emph{{Axion assisted Schwinger effect}}, \href{https://doi.org/10.1007/JHEP05(2021)001}{\emph{JHEP} {\bfseries 05} (2021) 001} [\href{https://arxiv.org/abs/2101.05192}{{\ttfamily 2101.05192}}].

\bibitem{vonEckardstein:2024tix}
R.~von Eckardstein, K.~Schmitz and O.~Sobol, \emph{{On the Schwinger effect during axion inflation}}, \href{https://doi.org/10.1007/JHEP02(2025)096}{\emph{JHEP} {\bfseries 02} (2025) 096} [\href{https://arxiv.org/abs/2408.16538}{{\ttfamily 2408.16538}}].

\bibitem{Iarygina:2025ncl}
O.~Iarygina, E.~I. Sfakianakis and A.~Brandenburg, \emph{{Schwinger effect in axion inflation on a lattice}},  \href{https://arxiv.org/abs/2506.20538}{{\ttfamily 2506.20538}}.

\bibitem{Adshead:2015pva}
P.~Adshead, J.~T. Giblin, T.~R. Scully and E.~I. Sfakianakis, \emph{{Gauge-preheating and the end of axion inflation}}, \href{https://doi.org/10.1088/1475-7516/2015/12/034}{\emph{JCAP} {\bfseries 12} (2015) 034} [\href{https://arxiv.org/abs/1502.06506}{{\ttfamily 1502.06506}}].

\bibitem{Adshead:2023mvt}
P.~Adshead, J.~T. Giblin, R.~Grutkoski and Z.~J. Weiner, \emph{{Gauge preheating with full general relativity}}, \href{https://doi.org/10.1088/1475-7516/2024/03/017}{\emph{JCAP} {\bfseries 03} (2024) 017} [\href{https://arxiv.org/abs/2311.01504}{{\ttfamily 2311.01504}}].

\bibitem{Caravano:2021bfn}
A.~Caravano, E.~Komatsu, K.~D. Lozanov and J.~Weller, \emph{{Lattice simulations of Abelian gauge fields coupled to axions during inflation}}, \href{https://doi.org/10.1103/PhysRevD.105.123530}{\emph{Phys. Rev. D} {\bfseries 105} (2022) 123530} [\href{https://arxiv.org/abs/2110.10695}{{\ttfamily 2110.10695}}].

\bibitem{Figueroa:2023oxc}
D.~G. Figueroa, J.~Lizarraga, A.~Urio and J.~Urrestilla, \emph{{Strong Backreaction Regime in Axion Inflation}}, \href{https://doi.org/10.1103/PhysRevLett.131.151003}{\emph{Phys. Rev. Lett.} {\bfseries 131} (2023) 151003} [\href{https://arxiv.org/abs/2303.17436}{{\ttfamily 2303.17436}}].

\bibitem{Sharma:2024nfu}
R.~Sharma, A.~Brandenburg, K.~Subramanian and A.~Vikman, \emph{{Lattice simulations of axion-U(1) inflation: gravitational waves, magnetic fields, and scalar statistics}}, \href{https://doi.org/10.1088/1475-7516/2025/05/079}{\emph{JCAP} {\bfseries 05} (2025) 079} [\href{https://arxiv.org/abs/2411.04854}{{\ttfamily 2411.04854}}].

\bibitem{Maleknejad:2020pec}
A.~Maleknejad, \emph{{Chiral anomaly in SU(2)$_{R}$-axion inflation and the new prediction for particle cosmology}}, \href{https://doi.org/10.1007/JHEP06(2021)113}{\emph{JHEP} {\bfseries 21} (2020) 113} [\href{https://arxiv.org/abs/2103.14611}{{\ttfamily 2103.14611}}].

\bibitem{Maleknejad:2020yys}
A.~Maleknejad, \emph{{SU(2)R and its axion in cosmology: A common origin for inflation, cold sterile neutrinos, and baryogenesis}}, \href{https://doi.org/10.1103/PhysRevD.104.083518}{\emph{Phys. Rev. D} {\bfseries 104} (2021) 083518} [\href{https://arxiv.org/abs/2012.11516}{{\ttfamily 2012.11516}}].

\bibitem{Alexander:2004us}
S.~H.-S. Alexander, M.~E. Peskin and M.~M. Sheikh-Jabbari, \emph{{Leptogenesis from gravity waves in models of inflation}}, \href{https://doi.org/10.1103/PhysRevLett.96.081301}{\emph{Phys. Rev. Lett.} {\bfseries 96} (2006) 081301} [\href{https://arxiv.org/abs/hep-th/0403069}{{\ttfamily hep-th/0403069}}].

\bibitem{Maleknejad:2014wsa}
A.~Maleknejad, \emph{{Chiral Gravity Waves and Leptogenesis in Inflationary Models with non-Abelian Gauge Fields}}, \href{https://doi.org/10.1103/PhysRevD.90.023542}{\emph{Phys. Rev. D} {\bfseries 90} (2014) 023542} [\href{https://arxiv.org/abs/1401.7628}{{\ttfamily 1401.7628}}].

\bibitem{Maleknejad:2016dci}
A.~Maleknejad, \emph{{Gravitational leptogenesis in axion inflation with SU(2) gauge field}}, \href{https://doi.org/10.1088/1475-7516/2016/12/027}{\emph{JCAP} {\bfseries 12} (2016) 027} [\href{https://arxiv.org/abs/1604.06520}{{\ttfamily 1604.06520}}].

\bibitem{Maleknejad:2024vvf}
A.~Maleknejad, \emph{{Gravitational ABJ Anomaly, Stochastic Matter Production, and Leptogenesis}},  \href{https://arxiv.org/abs/2412.09490}{{\ttfamily 2412.09490}}.

\bibitem{Maleknejad:2024ybn}
A.~Maleknejad and J.~Kopp, \emph{{Gravitational Wave-Induced Freeze-In of Fermionic Dark Matter}},  \href{https://arxiv.org/abs/2405.09723}{{\ttfamily 2405.09723}}.

\bibitem{Maleknejad:2024hoz}
A.~Maleknejad and J.~Kopp, \emph{{Weyl fermion creation by cosmological gravitational wave background at 1-loop}}, \href{https://doi.org/10.1007/JHEP01(2025)023}{\emph{JHEP} {\bfseries 01} (2025) 023} [\href{https://arxiv.org/abs/2406.01534}{{\ttfamily 2406.01534}}].

\bibitem{Garani:2024isu}
R.~Garani, M.~Redi and A.~Tesi, \emph{{Stochastic Dark Matter from Curvature Perturbations}}, \href{https://doi.org/10.1103/PhysRevLett.134.101005}{\emph{Phys. Rev. Lett.} {\bfseries 134} (2025) 101005} [\href{https://arxiv.org/abs/2408.15987}{{\ttfamily 2408.15987}}].

\bibitem{Garani:2025qnm}
R.~Garani, M.~Redi and A.~Tesi, \emph{{Particle production from inhomogeneities: general metric perturbations}}, \href{https://doi.org/10.1007/JHEP08(2025)037}{\emph{JHEP} {\bfseries 08} (2025) 037} [\href{https://arxiv.org/abs/2502.12249}{{\ttfamily 2502.12249}}].

\bibitem{Eroncel:2025qlk}
C.~Er{\"o}ncel, Y.~Gouttenoire, R.~Sato, G.~Servant and P.~Simakachorn, \emph{{A New Source for (QCD) Axion Dark Matter Production: Curvature-Induced}},  \href{https://arxiv.org/abs/2503.04880}{{\ttfamily 2503.04880}}.

\bibitem{Schwartz_2013}
M.~D. Schwartz, \emph{Quantum Field Theory and the Standard Model}. Cambridge University Press, 2013.

\bibitem{Fabbrichesi_2021}
M.~Fabbrichesi, E.~Gabrielli and G.~Lanfranchi, \emph{The Physics of the Dark Photon: A Primer}. Springer International Publishing, 2021, \href{https://doi.org/10.1007/978-3-030-62519-1}{10.1007/978-3-030-62519-1}.

\bibitem{Caputo:2021eaa}
A.~Caputo, A.~J. Millar, C.~A.~J. O'Hare and E.~Vitagliano, \emph{{Dark photon limits: A handbook}}, \href{https://doi.org/10.1103/PhysRevD.104.095029}{\emph{Phys. Rev. D} {\bfseries 104} (2021) 095029} [\href{https://arxiv.org/abs/2105.04565}{{\ttfamily 2105.04565}}].

\bibitem{Bogorad:2021uew}
Z.~Bogorad and N.~Toro, \emph{{Ultralight millicharged dark matter via misalignment}}, \href{https://doi.org/10.1007/JHEP07(2022)035}{\emph{JHEP} {\bfseries 07} (2022) 035} [\href{https://arxiv.org/abs/2112.11476}{{\ttfamily 2112.11476}}].

\bibitem{Maleknejad:2022gyf}
A.~Maleknejad and E.~McDonough, \emph{{Ultralight pion and superheavy baryon dark matter}}, \href{https://doi.org/10.1103/PhysRevD.106.095011}{\emph{Phys. Rev. D} {\bfseries 106} (2022) 095011} [\href{https://arxiv.org/abs/2205.12983}{{\ttfamily 2205.12983}}].

\bibitem{JonaLasinio1964}
G.~Jona-Lasinio, \emph{Relativistic field theories with symmetry-breaking solutions}, \href{https://doi.org/10.1007/BF02750573}{\emph{Nuovo Cimento} {\bfseries 34} (1964) 1790}.

\bibitem{ItzyksonZuber1980}
C.~Itzykson and J.-B. Zuber, \emph{Quantum Field Theory}. McGraw--Hill, New York, 1980.

\bibitem{Peskin:1995ev}
M.~E. Peskin and D.~V. Schroeder, \emph{{An Introduction to quantum field theory}}. Addison-Wesley, Reading, USA, 1995, \href{https://doi.org/10.1201/9780429503559}{10.1201/9780429503559}.

\bibitem{Kim:2003qp}
S.~P. Kim and D.~N. Page, \emph{{Schwinger pair production in electric and magnetic fields}}, \href{https://doi.org/10.1103/PhysRevD.73.065020}{\emph{Phys. Rev. D} {\bfseries 73} (2006) 065020} [\href{https://arxiv.org/abs/hep-th/0301132}{{\ttfamily hep-th/0301132}}].

\bibitem{Narozhny:1970uv}
N.~B. Narozhny and A.~I. Nikishov, \emph{{The Simplest processes in the pair creating electric field}}, {\emph{Sov. Phys. JETP} {\bfseries 38} (1974) 427}.

\bibitem{Dunne:2004nc}
G.~V. Dunne, \emph{{Heisenberg-Euler effective Lagrangians: Basics and extensions}}, \href{https://doi.org/10.1142/9789812775344_0014}{\emph{From fields to strings: Circumnavigating theoretical physics} {\bfseries 1} (2005) 445} [\href{https://arxiv.org/abs/hep-th/0406216}{{\ttfamily hep-th/0406216}}].

\bibitem{Weinberg:2008zzc}
S.~Weinberg, \emph{{Cosmology}}. Oxford University Press, 2008.

\bibitem{DeWitt:1964mxt}
B.~S. DeWitt, \emph{{Dynamical theory of groups and fields}}, {\emph{Conf. Proc. C} {\bfseries 630701} (1964) 585}.

\bibitem{Cohen1995}
L.~Cohen, \emph{Time-Frequency Analysis}. Prentice Hall, Englewood Cliffs, NJ, 1995.

\bibitem{Flandrin1999}
P.~Flandrin, \emph{Time--Frequency / Time--Scale Analysis}. Academic Press, 1999.

\bibitem{Feichtinger1998}
H.~G. Feichtinger and T.~Strohmer, \emph{Gabor analysis and algorithms: Theory and applications},  in \emph{Applied and Numerical Harmonic Analysis}.
\newblock Birkh{\"a}user, Boston, MA, 1998.
\newblock \href{https://doi.org/10.1007/978-1-4612-2016-1}{DOI}.

\bibitem{Oppenheim2009}
A.~V. Oppenheim, R.~W. Schafer and J.~R. Buck, \emph{Discrete-Time Signal Processing}. Prentice Hall, Upper Saddle River, NJ, 3rd~ed., 2009.

\bibitem{Mallat1999}
S.~Mallat, \emph{A Wavelet Tour of Signal Processing}. Academic Press, 2nd~ed., 1999.

\bibitem{Daubechies1992}
I.~Daubechies, \emph{Ten Lectures on Wavelets}. SIAM, 1992.

\bibitem{Raffelt:1996wa}
G.~G. Raffelt, \emph{{Stars as laboratories for fundamental physics}: {The astrophysics of neutrinos, axions, and other weakly interacting particles}}. University of Chicago Press, 5, 1996.

\bibitem{Band1993ApJ}
D.~Band, J.~Matteson, L.~Ford, B.~Schaefer, D.~Palmer, B.~Teegarden et~al., \emph{Batse observations of gamma-ray burst spectra. i. spectral diversity}, \href{https://doi.org/10.1086/172995}{\emph{The Astrophysical Journal} {\bfseries 413} (1993) 281}.

\bibitem{Davidson:2000hf}
S.~Davidson, S.~Hannestad and G.~Raffelt, \emph{{Updated bounds on millicharged particles}}, \href{https://doi.org/10.1088/1126-6708/2000/05/003}{\emph{JHEP} {\bfseries 05} (2000) 003} [\href{https://arxiv.org/abs/hep-ph/0001179}{{\ttfamily hep-ph/0001179}}].

\bibitem{Cora2019}
C.~Dvorkin, T.~Lin and K.~Schutz, \emph{Making dark matter out of light: Freeze-in from plasma effects}, \href{https://doi.org/10.1103/PhysRevD.99.115009}{\emph{Phys. Rev. D} {\bfseries 99} (2019) 115009}.

\bibitem{Fedderke_2021}
M.~A. Fedderke, P.~W. Graham, D.~F. Jackson~Kimball and S.~Kalia, \emph{Search for dark-photon dark matter in the supermag geomagnetic field dataset}, \href{https://doi.org/10.1103/PhysRevD.104.095032}{\emph{Phys. Rev. D} {\bfseries 104} (2021) 095032}.

\bibitem{Cline:2024wja}
J.~M. Cline and G.~Herrera, \emph{{Plausible constraints and inflationary production for dark photons}}, \href{https://doi.org/10.1103/PhysRevD.112.035023}{\emph{Phys. Rev. D} {\bfseries 112} (2025) 035023} [\href{https://arxiv.org/abs/2409.13818}{{\ttfamily 2409.13818}}].

\bibitem{Caputo:2025avc}
A.~Caputo, J.~Park and S.~Yun, \emph{{The Heavy Dark Photon Handbook: Cosmological and Astrophysical Bounds}},  \href{https://arxiv.org/abs/2511.15785}{{\ttfamily 2511.15785}}.

\bibitem{Lozanov:2019jxc}
K.~D. Lozanov, \emph{{Lectures on Reheating after Inflation}},  \href{https://arxiv.org/abs/1907.04402}{{\ttfamily 1907.04402}}.

\bibitem{Maleknejad:2023nyh}
A.~Maleknejad, \emph{{Photon chiral memory effect stored on celestial sphere}}, \href{https://doi.org/10.1007/JHEP06(2023)193}{\emph{JHEP} {\bfseries 06} (2023) 193} [\href{https://arxiv.org/abs/2304.05381}{{\ttfamily 2304.05381}}].

\bibitem{Kobayashi:2014zza}
T.~Kobayashi and N.~Afshordi, \emph{{Schwinger Effect in 4D de Sitter Space and Constraints on Magnetogenesis in the Early Universe}}, \href{https://doi.org/10.1007/JHEP10(2014)166}{\emph{JHEP} {\bfseries 10} (2014) 166} [\href{https://arxiv.org/abs/1408.4141}{{\ttfamily 1408.4141}}].

\bibitem{Bavarsad:2017oyv}
E.~Bavarsad, S.~P. Kim, C.~Stahl and S.-S. Xue, \emph{{Effect of a magnetic field on Schwinger mechanism in de Sitter spacetime}}, \href{https://doi.org/10.1103/PhysRevD.97.025017}{\emph{Phys. Rev. D} {\bfseries 97} (2018) 025017} [\href{https://arxiv.org/abs/1707.03975}{{\ttfamily 1707.03975}}].

\bibitem{Maleknejad:2011sq}
A.~Maleknejad and M.~M. Sheikh-Jabbari, \emph{{Non-Abelian Gauge Field Inflation}}, \href{https://doi.org/10.1103/PhysRevD.84.043515}{\emph{Phys. Rev. D} {\bfseries 84} (2011) 043515} [\href{https://arxiv.org/abs/1102.1932}{{\ttfamily 1102.1932}}].

\bibitem{Maleknejad:2011jw}
A.~Maleknejad and M.~M. Sheikh-Jabbari, \emph{{Gauge-flation: Inflation From Non-Abelian Gauge Fields}}, \href{https://doi.org/10.1016/j.physletb.2013.05.001}{\emph{Phys. Lett. B} {\bfseries 723} (2013) 224} [\href{https://arxiv.org/abs/1102.1513}{{\ttfamily 1102.1513}}].

\end{thebibliography}
\end{document}